\documentclass[letterpaper]{article} 
\usepackage{aaai23}  
\usepackage{times}  
\usepackage{helvet}  
\usepackage{courier}  
\usepackage[hyphens]{url}  
\usepackage{graphicx} 
\usepackage{natbib}  
\usepackage{caption} 
\usepackage{algorithm}
\usepackage{algorithmic}

\usepackage{amsmath}
\usepackage{amssymb}
\usepackage{mathtools}
\usepackage{amsthm}
\usepackage{bm}
\usepackage{subfigure}
\usepackage{dblfloatfix} 
\def\inte{INTENT}
\def\bdrl{BACKDOORL}
\usepackage{newfloat}
\usepackage{listings}
\DeclareCaptionStyle{ruled}{labelfont=normalfont,labelsep=colon,strut=off} 
\floatstyle{ruled}
\newfloat{listing}{tb}{lst}{}
\floatname{listing}{Listing}
\title{Backdoor Attacks on Multiagent Collaborative Systems}
\author{
    Shuo Chen, Yue Qiu, Jie Zhang\\
}
\affiliations{
    School of Computer Science and Engineering\\
    Nanyang Technological University\\
    50 Nanyang Ave, Singapore, 639798\\
    \{schen,yue.qiu,zhangj\}@ntu.edu.sg
}

\begin{document}

\maketitle

\begin{abstract}
Backdoor attacks on reinforcement learning implant a backdoor in a victim agent's policy. Once the victim observes the trigger signal, it will switch to the abnormal mode and fail its task. 
Most of the attacks assume the adversary can arbitrarily modify the victim's observations, which may not be practical. One work proposes to let one adversary agent use its actions to affect its opponent in two-agent competitive games, so that the opponent quickly fails after observing certain trigger actions.
However, in multiagent collaborative systems, agents may not always be able to observe others.
When and how much the adversary agent can affect others are uncertain, and we want the adversary agent to trigger others for as few times as possible. 
To solve this problem, we first design a novel training framework to produce auxiliary rewards that measure the extent to which the other agents' observations being affected. Then we use the auxiliary rewards to train a trigger policy which enables the adversary agent to efficiently affect the others' observations. Given these affected observations, we further train the other agents to perform abnormally. 
Extensive experiments demonstrate that the proposed method enables the adversary agent to lure the others into the abnormal mode with only a few actions.
\end{abstract}

\section{Introduction}
Recent studies \cite{yangIntentional,kiourtiTrojDRL,ashcraftPoisoning,yu2022temporal,chen2022marnet} have shown that the deep reinforcement learning (DRL) is vulnerable to backdoor attacks. 
Intuitively, backdoor attacks embed both benign and abnormal modes in a victim agent's policy. The victim agent usually exhibits the benign behaviors and completes its task successfully. However,
once the victim agent observes the trigger signal, it will switch to the abnormal mode and fail its task.
Most of the existing attacks assume the adversary has the ability to modify the victim agent's observations arbitrarily. For instance, it directly modifies the environment being observed by the victim or manipulates the victim's sensors. Because this assumption can be strong in practical scenarios,  
one work \cite{wangBACKDOORL} proposes to let one adversary agent use its actions to affect its opponent in two-agent competitive games. Once the opponent observes certain trigger actions performed by the adversary agent, it  fails quickly.

While being more practical, the method proposed in \cite{wangBACKDOORL} may not be directly applied to multiagent collaborative systems. The reason is that the environment of multiagent collaborative systems is normally partially observable. This means agents cannot always observe others. Even the adversary agent performs trigger actions, there is no guarantee that those actions being observed. Therefore, to implant a backdoor in the collaborative team policy, the key problem is to learn when and how much the adversary agent can affect its teammates. Moreover, we want the adversary agent to trigger its teammates for as few times as possible to minimize the risk of being detected. 

To solve this problem, we first learn to measure the influence of the adversary agent on the teammates' observations. Specifically,
we maintain two neural networks as the observation encoders and fix one encoder's parameters while leaving the other trainable. Given a sampled episode, both observation encoders transform each teammate observation into an encoding vector. Only when the episode is benign, i.e., the adversary agent does not perform trigger actions, we train the trainable encoder to minimize the distance between the vectors output by the two encoders. This results in that if an observation is quite different from those of benign episodes, the distance between its encoding vectors will be large. 
Therefore, when we generate an abnormal episode by letting the adversary agent perform trigger actions, the encoding vector distances can serve as the auxiliary rewards that measure the difference between the observations from the benign and abnormal episodes. Based on the auxiliary rewards, we further train a trigger policy. This policy instructs the adversary agent to selectively trigger the teammates so that their observations are quite different from those in benign episodes.  
Given these affected observations, we train the teammates to behave maliciously. Note that agents select actions based on their observations. The observation difference between the benign and abnormal episodes ensures that the malicious behavior training will not affect the benign behavior. This finally allows us to embed both benign and abnormal modes in the collaborative team policy. 

To summarize, the contributions of this paper are threefold: 1) to the best of our knowledge, this is the first work that uses the adversary agent to trigger its teammates' backdoor in multiagent collaborative systems. 2) We explicitly measure the extent to which the teammates' observations being affected by the adversary agent. Based on the measurements, we propose to learn a trigger policy which enables the adversary agent to lure the teammates into the abnormal mode by only a few actions. 
3) We have conducted extensive experiments to demonstrate the effectiveness of the proposed method. When compared to the state-of-the-art methods, the team policy trained by our method not only achieves a higher task performance when without the trigger actions, but also compromises the collaboration more severely when the backdoor is triggered.

\section{Related Work}
\label{related}
The backdoor attack was first identified in \cite{gu2017badnets}, where the adversary trains an image classification network that has good performance on normal training and validation samples while behaving poorly when given specific inputs chosen by the adversary. Following this work, \cite{chen2017targeted, liu2017trojaning} have achieved similar attacks with weaker assumptions regarding the capabilities of the adversary. Besides the image classification systems, researchers also show that backdoor attacks can  be launched in natural language processing systems \cite{chen2021badnl}, generative adversarial networks \cite{salem2020baaan}, and federated learning systems \cite{bhagoji2019analyzing, xie2019dba, wang2020attack}.

Recent studies have demonstrated that the adversary can implant a backdoor in DRL policies by manipulating the policy inputs and rewards received by the agent during the training. In the work of \cite{kiourtiTrojDRL}, the agent receives a positive reward if its policy outputs an action desired by the adversary when triggered, otherwise the agent receives a negative reward. However, this reward manipulation only teaches the agent to deviate from its normal behavior when triggered. This does not mean the triggered actions will fail the task effectively.
In comparison, \cite{yangIntentional} do not specify the desired actions. Instead, they change the rewards received by the agent to the opposite of the rewards returned by the environment after the trigger signal is sent. Based on the manipulated rewards, the agent can learn how to fail the task after seeing the trigger signal. \cite{ashcraftPoisoning,yu2022temporal} adopt a similar training procedure to that of \cite{yangIntentional} with the consideration of novel types of trigger. \cite{ashcraftPoisoning} propose in-distribution triggers and 
\cite{yu2022temporal} study temporal-pattern triggers. 
The work \cite{chen2022marnet} extends the attacks on single-agent systems to multiagent collaborative systems. To create poisoned training data, they directly modify the environment and let the agents, which observe the environmental change, select the worst actions suggested by an expert model.

All the above attacks on DRL assume a strong attacker capability, i.e., the adversary can directly modify the victim agents' observations. To eliminate this assumption, \cite{wangBACKDOORL} propose to affect the opponent's observations by the agent interaction in two-agent competitive games. Although they achieve to trigger the opponent's backdoor by the adversary agent's actions, they do not consider the partially observable cases where the trigger actions may not be observed. This issue can either damage the training of backdoor policies or require the adversary agent to perform a lot of trigger actions, which increases the possibility of being detected. In comparison, this work explicitly considers this issue and trains a trigger policy to determine when the adversary agent should perform trigger actions. Moreover, the backdoor policy training in multiagent collaborative systems is harder than that in two-agent competitive games because the collaborative agents' behavior may affect one another during the training. Besides \cite{wangBACKDOORL}, \cite{wangTransport} also use the adversary agent's actions to trigger the backdoor. However, this attack is designed for learning-based traffic congestion
control systems and thus, heavily relies on the principles of traffic systems.


\section{Background}
We represent the teamwork in multiagent collaborative systems as a decentralized partially observable Markov decision process (Dec-POMDP) \cite{oliehoek2016concise} consisting of a tuple $\langle N, {S}, \{A_i\}, {T}, {R}, \Omega, O, \gamma \rangle$. $N\equiv\{1,...,n\}$ is the set of team agents. $s\in{S}$ describes the state of all team agents and the external environment. $ A_i$ is the set of actions of an agent $i\in N$ and $A=\times_{i=1}^{n}A_i$ is the set of joint actions. 
In every state $s$, each agent $i\in N$ gets an observation $o_i\in \Omega$ based on the observation function $O:S\times A\rightarrow\Omega$. Each agent $i$ makes its decision based on its own policy $\pi_i:\Omega\times A_i\rightarrow [0,1]$ that specifies the probability $\pi_i(a_i|o_i)$ of taking action $a_i$ when given the observation $o_i$.
The actions chosen by all agents form a joint action $\mathbf{a}=\{a_1,...,a_n\}\in A$. The joint action $\mathbf{a}$ causes the transition from the state $s$ to the next state $s'$ according to the state transition function ${{T:S\times A\times S\rightarrow[0,1]}}$ that specifies the transition probability ${{P}}({s}'|{s}, \mathbf{a}) \ \forall \ {s}'\in {{S}}$. After the transition, all team agents get a team reward $r(s, \mathbf{a})$ specified by the reward function $R:S\times A \rightarrow \mathcal{R}$. $\gamma$ is the discount factor for the future reward where $0\leq \gamma < 1$.

Since agents' actions are conditioned on their observations and the observations are conditioned on states, we can use a 
joint team policy $\bm{\pi}:S\times A\rightarrow [0,1]$ to represent the probability $\bm{\pi}(\mathbf{a}|s)$ of taking joint action $\mathbf{a}$ in state $s$. 
The Q-value of a joint action $\mathbf{a}$ in state $s$ is,
\begin{equation}
	\label{jointQ}
	Q^{\bm{\pi}}(s,\mathbf{a})=r(s,\mathbf{a})+\gamma\sum_{s'\in S}{P(s'|s,\mathbf{a}) V^{\bm{\pi}}(s')}.
\end{equation}
It represents the expected team return when the team takes the joint action $\mathbf{a}$ in state $s$ and then follows the policy $\bm{\pi}$.

\subsection{Multiagent Reinforcement Learning}
For multiagent collaborative systems, multiagent reinforcement learning (MARL) is a popular way to learn the joint team policy $\bm{\pi}$. 
Since team agents make decisions based on observations, i.e., partial state information, it could be hard to find a satisfied team policy. To solve this problem, it is a common practice to adopt the centralized training with decentralized execution (CTDE) framework \cite{oliehoek2008optimal,hernandez2019survey}. In this framework, the centralized training uses the global information, i.e., full state information, the history of states and actions, while agents still only get observations during the decentralized execution stage. 

In this work, we 
use one of the most cited MARL algorithms, i.e., QMIX \cite{qmix}, to train both benign and abnormal behaviors. In QMIX, 
given an observation $o_i$ and action $a_i$, agent $i$'s policy network $Q_i$ outputs a local Q-value estimation $Q_i(o_i, a_i)$. QMIX further aggregates the local Q-value estimation of all agents into a total Q-value estimation $Q_{total}(\bm{o},\bm{a}, s)$ by a mixing network that conditions on the global state $s$. 
Moreover, QMIX forces the mixing network to have positive weights. This ensures the monotonicity relationship between $Q_{total}(\bm{o},\bm{a}, s)$ and $Q_i(o_i, a_i)$,
\begin{equation}
\frac{\partial Q_{total}(\bm{o},\bm{a}, s)}{\partial Q_i(o_i, a_i)}\geq 0, \forall a_i \in A_i.	
\end{equation}
This relationship means 
when QMIX finds the optimal total Q-value $Q_{total}(\bm{o},\bm{a}, s)$, the action with the maximum $Q_i(o_i, a_i)$ is the optimal action of agent $i$. The learning of the optimal total Q-value follows the traditional Q-learning procedure, i.e., minimizing the following loss,
\begin{equation}
	\label{qmix_loss}
	\mathcal{L}=\sum_{k=1}^{b}[(y_{total}^k-Q_{total}(\bm{o}, \bm{a}, s))^2],
\end{equation}
where $b$ is the number of transitions sampled from the replay buffer, $y_{total}^k=r(s,\mathbf{a})+\gamma \text{max}_{\bm{a}'}\bar{Q}_{total}(\bm{o}', \bm{a}', s')$, and $\bar{Q}$ means the total Q-value of the next state-action pair is output by a target network as in DQN \cite{mnih2015human}.

\subsection{Threat Model}
\label{threat}
For the backdoor attack, we consider two parties, i.e., the user and the adversary. Because the team policy training often consumes a lot of resources, the user wants to outsource the training task to the cloud platforms that provide the training service. The adversary can serve as the contractor of the cloud platforms. During the model training, the adversary can manipulate the global team rewards received by agents and the actions chosen by agents. However, the adversary cannot directly access the model parameters, i.e., the trained models are also black-box to the adversary. Also, the adversary cannot arbitrarily modify the observations of agents.

After the trained models are deployed in the user side, the adversary can only use the adversary agent to trigger the backdoor. Note that we assume there is only one adversary agent in the team. This assumption makes the attack more practical than having multiple adversary agents. The adversary agent usually cooperates with its teammates. To let itself start to trigger the backdoor,
the adversary agent needs an 1-bit external signal. This signal can be sent either by the adversary through communication or by a timing program inside the adversary agent. 
How to generate this signal depends on the specific domains. Actually, the work \cite{wangBACKDOORL} also needs such a signal to explicitly tell the adversary agent whether it can perform trigger actions. If the adversary agent is always allowed to perform trigger actions, its opponent or teammates will always fail. In this case, the user will not deploy the trained models. 

\section{Backdoor Attack on Teamwork}
In this section, we first explain how to produce the auxiliary rewards that measure how much the observations have been affected by the trigger actions. Then we present the details about the trigger policy training based on the auxiliary rewards. Finally, we describe the overall procedure of the backdoor attack on multiagent collaborative systems.

\subsection{Auxiliary Reward Generation}
We aim to let the adversary agent perform trigger actions to trigger its teammates' backdoor. However, in a partially observable environment, the trigger actions may not be observed by the teammates and thus, may not take effect. Although the adversary agent may perform the trigger action for many times to increase the probability of being observed, this also increases the risk of being detected. Therefore, the adversary agent needs to learn how much the trigger actions can affect its teammates' observations so that it can choose to perform the trigger action only when it has a big impact. 

An intuitive way is to let the adversary agent try both choices, i.e., performing a trigger action or performing a benign action, in a same state and compare the next observations resulting from the two choices. However, this requires simulators or real environments to rollback to the previous state after the adversary agent tries one of the choices, which either significantly increases the simulation complexity or is infeasible in reality. Moreover, this comparison does not consider the long-term effects of trigger actions. That is, when the adversary agent performs a trigger action in a certain step, the observations in several steps later may also be affected because the state distribution has been changed by the trigger action.
Therefore, to evaluate the overall effects of trigger actions, we should let the adversary agent randomly perform trigger actions to generate full training episodes. The episodes without trigger actions are benign episodes while the episodes with at least one trigger action are abnormal episodes. Our goal is to find an efficient way to compare the observations of the two kinds of episodes. 

However, there are many training episodes, so it is impractical to check every pair of benign and abnormal episodes. 
Inspired by the exploration work \cite{burda2018exploration}, we propose to use the random network distillation (RND) module to memorize the observations seen during benign episodes. Then given an observation from abnormal episodes, the RND module can determine how similar this observation is to the memorized observations. 
Specifically, the RND module consists of two observation encoders $\mathcal{E}:\Omega\rightarrow \mathcal{R}^m$. Each encoder takes an observation as input and outputs an $m$-dimensional vector. We randomly initialize the two encoders and fix the parameters of one network while leaving another trainable. We denote the fixed one as $\bar{\mathcal{E}}(\cdot)$ and the trainable one as $\mathcal{E}(\cdot;\theta)$ where $\theta$ represents the trainable parameters. 
To let the RND module memorize the observations from benign episodes, we simply input each observation $o$ into the encoders and obtain two encoding vectors, i.e., $\bar{\mathcal{E}}(o)$ and $\mathcal{E}(o;\theta)$, respectively. Then we minimize the L2-norm distance between the encoding vectors,
\begin{equation}
	\label{ob_loss}
	\mathcal{L}_{\theta}=\sum_{k=1}^{b}\|\mathcal{E}(o_k;\theta)-\bar{\mathcal{E}}(o_k)\|_2,
\end{equation}
where $b$ is the number of observations sampled from benign episodes. This minimization serves as the memorization process. It ensures that the trained $\mathcal{E}(\cdot;\theta)$ will output a vector close to that output by $\bar{\mathcal{E}}(\cdot)$ when given an observation from benign episodes. Then given an observation $o$ from abnormal episodes, the encoding vector distance can measure the extent to which $o$ deviates from the observations in benign episodes. If none of the observations in benign episodes is similar to $o$, there will be a high probability that the encoding vector distance is large. In the next subsection, we will use this distance as the auxiliary reward, i.e.,
\begin{equation}
	\label{robs}
	r_{obs}(o)=\|\mathcal{E}(o;\theta)-\bar{\mathcal{E}}(o)\|_2,
\end{equation}
to train the trigger policy.

\begin{figure*}[t]
	\centering
	\includegraphics[width=0.8\textwidth]{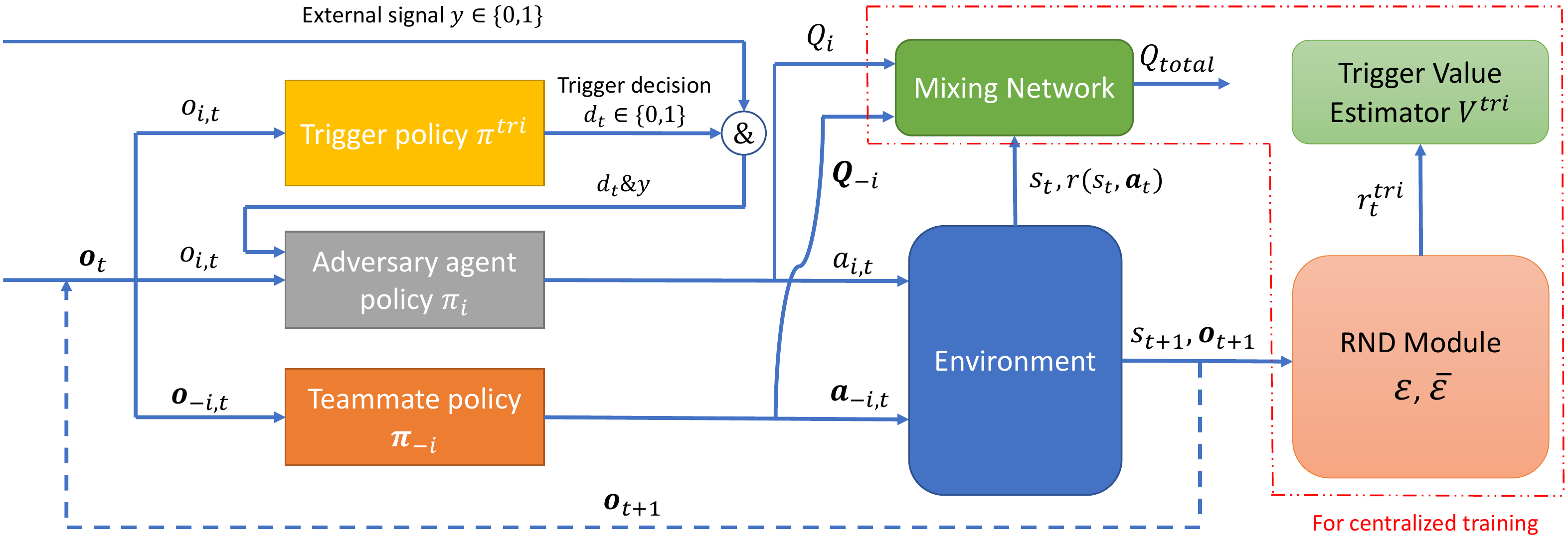} 
	\caption{The overall attack framework.}
	\label{fig_framework}
\end{figure*}

\subsection{Trigger Policy Training}
An ideal attack is to maximize the influence on teammates' observations while performing trigger actions for as few times as possible. To this end,
we propose to learn a trigger policy.
Specifically, we let $i$ denote the index of the adversary agent and initialize the trigger policy as a neural network $\pi^{tri}:\Omega_i \times \{0,1\} \rightarrow [0, 1]$. Given the adversary agent $i$'s observation $o_i$, the trigger policy specifies the probability $\pi^{tri}(d|o_i)$ of the trigger decision $d$ which is either $0$ or $1$. If $d$ is $1$, it means the adversary agent will perform the trigger action. Otherwise, it will perform the benign action. We formalize the learning objective as,
\begin{align}
	\label{tri_obj}
	& \max_{d_t\sim \pi^{tri}} \mathbb{E}_{(\bm{o}, \bm{a})\sim\tau} \sum_{t=0}^T \gamma^t r^{tri}_t \notag\\
	& \text{with}  \:\: r^{tri}_t=[\frac{1}{N-1}\sum_{k\in N, k\neq i}r_{obs}(o_{k,t+1})]-\alpha \cdot d_t,
\end{align}
where $\tau$ is the trajectory distribution induced by the joint team policy, $d_t$ means the trigger decision in the time step $t$, $o_{k,t+1}$ is the observation received by the teammate $k$ in the time step $t+1$, $r_{obs}(o_{k,t+1})$ is computed based on Equation (\ref{robs}), and $\alpha$ is a value used to balance the influence on teammates' observations and the number of trigger times. We compute $\alpha$ as the mean value of $r_{obs}$ across the sampled episodes. Note that we use the teammates' observations in the time step $t+1$ to compute $r_t^{tri}$ because the trigger decision cannot affect the observations in the same time step. $r_{obs}(o_{k,t+1})$ can represent the immediate influence of $d_t$ on $o_{k,t+1}$. 
In addition, the learning objective is to maximize the discounted  cumulative reward $\sum_{t=0}^T \gamma^t r^{tri}_t$, which captures the long-term influence of trigger decisions.

Given a batch of abnormal episodes, we can compute the $\{r_t^{tri}|t=0,\cdots,T\}$ based on their observations and the RND module. To achieve the learning objective, we train the trigger policy by the REINFORCE method \cite{sutton2018reinforcement}. 
Let $\phi$ denote the parameters of $\pi^{tri}$, we minimize the following loss,
\begin{align}
	\label{tri_pi_loss}
	&\mathcal{L}_{\phi}=\mathbb{E}_{(\bm{o_t}, \bm{a_t})\sim\tau, d_t\sim \pi^{tri}} \big[-A_t\cdot  log\pi^{tri}(d_t|o_{i,t};\phi)\big] \notag\\
	&\text{with  } A_t=\sum_{k=0}^{T-t} \gamma^k r^{tri}_{t+k}-V^{tri}(s_t;\varphi),
\end{align}
where $V^{tri}(s_t;\varphi)$ is the trigger value estimator parameterized by $\varphi$. It estimates the expected influence of trigger decisions starting from the state $s_t$. We use it to reduce the variance of gradient estimation. To learn the trigger value estimator, we minimize the following loss,
\begin{equation}
	\label{tri_v_loss}
	\mathcal{L}_{\varphi}=\mathbb{E}_{(\bm{o_t}, \bm{a_t})\sim\tau, d_t\sim \pi^{tri}}\Big[\big(\sum_{k=0}^{T-t} \gamma^k r^{tri}_{t+k}-V^{tri}(s_t;\varphi)\big)^2\Big].
\end{equation}
In the next subsection, we will describe the overall framework that uses the RND module and trigger policy to implant a backdoor in the team policy.

\subsection{Overall Attack Framework}
The overall attack framework is as shown in Figure \ref{fig_framework}. Note that $-i$ denotes the teammates of the adversary agent $i$, and $\&$ denotes the \emph{AND} operation between binary variables.
Each teammate $k$ maintains a policy network $\pi_k$.
The policy of all teammates form the joint teammate policy $\bm{\pi}_{-i}=\{\pi_k|k\in N, k\neq i\}$. The adversary agent $i$ also maintains a policy network $\pi_i$ that has a similar architecture to that of $\pi_k$. The only difference between them is that $\pi_i$ accepts an additional binary input, i.e., $d_t\& y$. This binary input leads to that the outputs of $\pi_i$ could be different when given the same observation and action, i.e., $\pi_i(o_{i,t}, a_{i,t}, 0)\neq \pi_i(o_{i,t}, a_{i,t}, 1)$. Therefore, the adversary agent determines whether it performs the trigger action based on this input. Besides $\pi_i$, the adversary also equips the adversary agent with the trigger policy $\pi^{tri}$. 

Another important thing in Figure \ref{fig_framework} is the external binary signal $y$. When generating an episode, we use this signal to decide whether this episode is a benign episode or an abnormal episode. When $y=0$, $d_t\& y=0$ no matter what $\pi^{tri}$ outputs. In this case, 
the adversary agent will perform benign actions and the resulting episode is a benign episode. When $y=1$, $d_t\& y=d_t$. In this case, 
if $d_t=1$, the adversary agent will perform a trigger action and the resulting episode is an abnormal episode. Note that $y$ only decides whether the adversary agent is allowed to perform trigger actions. It is $d_t$ that decides whether the adversary agent will perform a trigger action given $y=1$. For more discussion about the signal $y$, please refer to our threat model (Section \ref{threat}). 

\begin{algorithm}[!hb]
	\caption{The overall procedures for the backdoor attack}
	\label{alg_sample}
	\textbf{Input}: An environment $Env$, benign replay buffer $\mathcal{B}_b$, abnormal replay buffer $\mathcal{B}_a$, agent policies $\pi_i$ and $\bm{\pi}_{-i}$, trigger policy $\pi^{tri}$, training step limit $T_l$, benign training period $T_b$.
	
	\begin{algorithmic}[1] 
		\STATE Let $t_{total}=0$.
		\WHILE{$t_{total}<T_l$}
		\IF {$t_{total}<T_b$}
		\STATE Let $y=0$.
		\ELSE
		\STATE Let $y$ be $0$ or $1$ with equal probability.
		\ENDIF
		\STATE Let $t=0$, reset $Env$, and get $\bm{o}_t$
		\WHILE{$Env$ is not terminated}
		\STATE $d_t\leftarrow \pi^{tri}(o_{i,t})$
		\STATE $a_{i,t}\leftarrow \pi_i(o_{i,t},d_t\& y)$; $\bm{a}_{-i,t}\leftarrow \bm{\pi}_{-i}(o_{i,t})$
		\STATE $s_{t+1}, \bm{o}_{t+1}, r_t\leftarrow Env(a_{i,t},\bm{a}_{-i,t})$; $t\leftarrow t+1$.
		\ENDWHILE
		\STATE $t_{total}\leftarrow t_{total}+t$; Episode $e\leftarrow \{(s_t,\bm{o}_t,\bm{a}_t, d_t, r_t)\}$
		\IF {$\sum d_t\&y ==0$}
		\STATE $\mathcal{B}_b \leftarrow \mathcal{B}_b \cup e$
		\ELSE
		\STATE $\mathcal{B}_a \leftarrow \mathcal{B}_a \cup e$
		\ENDIF
		\STATE \emph{Training\_Function}($\mathcal{B}_b$, $\mathcal{B}_a$, $\pi_i$, $\bm{\pi}_{-i}$, $\pi^{tri}$)
		\ENDWHILE
	\end{algorithmic}
\end{algorithm}

\begin{algorithm}[!tb]
	\caption{\emph{Training\_Function}}
	\label{alg_train}
	\textbf{Input}: $\mathcal{B}_b$, $\mathcal{B}_a$, $\pi_i$, $\bm{\pi}_{-i}$, $\pi^{tri}$, mixing network $\mathcal{M}$, estimator $V^{tri}$, $RND$ module, batch size $b$.
	\begin{algorithmic}[1] 
		\IF {the size of $\mathcal{B}_b$ is larger than $b$} 
		\STATE Sample benign episodes $\{(s_t,\bm{o}_t,\bm{a}_t, d_t, r_t)\}$ 
		\STATE $\{Q_i\}\leftarrow \pi_i$,  $\{\bm{Q}_{-i}\}\leftarrow \bm{\pi}_{-i}$
		\STATE $\{Q_{total}\}\leftarrow  \{\mathcal{M}(Q_i, \bm{Q}_{-i})\}$  
		\STATE Minimize $\mathcal{L}$ in Equation (\ref{qmix_loss}).
		\STATE $\{(\mathcal{E}(o;\theta),\bar{\mathcal{E}}(o))\} \leftarrow RND$
		\STATE Minimize $\mathcal{L}_{\theta}$ in Equation (\ref{ob_loss}).
		\ENDIF
		
		\IF {the size of $\mathcal{B}_a$ is larger than $b$}
		\STATE Sample abnormal episodes $\{(s_t,\bm{o}_t,\bm{a}_t, d_t, r_t)\}$ 
		\STATE $\{Q_i\}\leftarrow \pi_i$,  $\{\bm{Q}_{-i}\}\leftarrow \bm{\pi}_{-i}$
		\STATE $\{Q_{total}\}\leftarrow  \{\mathcal{M}(Q_i, \bm{Q}_{-i})\}$
		\STATE Manipulate the rewards $\{r_t\}$.  
		\STATE Minimize $\mathcal{L}$ in Equation (\ref{qmix_loss}).
		\STATE$\{\pi^{tri}(d_t|o_{i,t})\}\leftarrow \pi^{tri}$
		\STATE $\{(\mathcal{E}(o;\theta),\bar{\mathcal{E}}(o))\} \leftarrow RND$
		\STATE$\{r_{obs}(o_k)\}\leftarrow \{(\mathcal{E}(o;\theta),\bar{\mathcal{E}}(o))\}$ \textit{// Equation (\ref{robs})} 
		\STATE $\{r_t^{tri}\}\leftarrow (\{r_{obs}(o_k)\}, \{d_t\})$ \textit{// Equation (\ref{tri_obj})} 
		\STATE Minimize $\mathcal{L}_{\phi}$ in Equation (\ref{tri_pi_loss}) and $\mathcal{L}_{\varphi}$ in Equation (\ref{tri_v_loss}).
		\ENDIF
	\end{algorithmic}
\end{algorithm}

We summarize the overall attack procedures as Algorithm \ref{alg_sample}. Note that we use two buffers in our algorithm to balance the percentage of abnormal episodes during training. The buffer $\mathcal{B}_b$ stores benign episodes while the buffer $\mathcal{B}_a$ stores abnormal episodes. In line $3\sim7$, we set the external signal $y$ based on the parameter $T_b$ which specifies the length of our benign training period. During this period, we always set $y$ as $0$. Therefore, we only sample benign episodes and thus, only train benign behavior and the RND module. This period can help the RND module to perform well when encountering abnormal episodes later. After the benign training period, we set $y$ as $0$ or $1$ with equal probability. In line $8\sim14$, we generate training episodes by interacting with the environment. 
In line $15\sim19$, we insert the generated episode into the corresponding buffer. If the sum of $\{d_t\&y\}$ is $0$ (line $15$), it means the adversary agent never performs trigger actions in this episode. Therefore, we insert this episode to the benign buffer $\mathcal{B}_b$. Otherwise, 
we insert it to the abnormal buffer $\mathcal{B}_a$. In line $20$, we conduct the training with sampled episodes. We wrap the training part in Algorithm \ref{alg_train}.

For Algorithm \ref{alg_train}, in line $1\sim8$, we train with benign episodes. The line $3\sim5$ follow the QMIX training procedures while in the line $6\sim7$, we train the RND module to minimize the encoding vector distance of the observations from benign episodes. In line $9\sim20$, we train with abnormal episodes. The line $11\sim14$ are similar to the line $3\sim5$. However, we add one reward manipulation step to enable the malicious behavior training. That is, if a trigger decision $d_t=1$ in the sampled episode, we set the rewards after the time step $t$, i.e., $\{r_j|j=t+1,\cdots,T\}$, to the opposite of their original values. By doing so, we can train the malicious behavior of $\pi_i$ and $\bm{\pi}_{-i}$ after the trigger action. In line $15\sim19$, we train the trigger policy $\pi^{tri}$ and its value estimator $V^{tri}$ based on Equation (\ref{robs}), (\ref{tri_obj}), (\ref{tri_pi_loss}), and (\ref{tri_v_loss}).

\section{Experiments}
In this section, we conduct extensive experiments to indicate the effectiveness of our attack. 
\subsubsection{Environment}
We conduct our experiments in the StarCraft Multi-Agent Challenge (SMAC) environment \cite{samvelyanSMAC}. This environment provides various maps each of which represents a unique task. 
In each map, there are two teams battling with each other. One team (i.e., the victim team we aim to attack) is open for training while the opponent team is controlled by built-in handcrafted heuristics.
In a task, an agent can only observe other alive agents that locate within its observation range. The reward function is shaped, i.e., the team will get a positive reward if its members hurt an opponent, kill an opponent, or win the game. 
If both teams are alive at the end of an episode, it is considered as a loss and the reward is $0$. 
We conduct our evaluations in three SMAC maps, i.e., \textit{3s\_vs\_3z}, \textit{8m}, and \textit{MMM}. In the \textit{3s\_vs\_3z} and \textit{8m} map, teams are homogeneous and the victim team has three and eight agents respectively. In the \textit{MMM} map, teams are heterogeneous and the victim team has three types of agents. Taking this into account, we have set the adversary agent as each of the three types to study whether the agent type would affect the attack. Therefore, we have investigated five attack scenarios, i.e., \textit{3s\_vs\_3z}, \textit{8m}, \textit{MMM0}, \textit{MMM3}, and \textit{MMM9} where the number after \textit{MMM} represents the index of the adversary agent inside the team.

\subsubsection{Baseline Methods}
We have implemented two baseline methods for comparison, i.e., \inte~\cite{yangIntentional} and \bdrl~\cite{wangBACKDOORL}. We choose \inte~as one baseline because it learns to fail the task based on reward manipulation, which makes its malicious behavior more effective than that of \cite{kiourtiTrojDRL}. Also, \inte~is the representative work of \cite{ashcraftPoisoning,yu2022temporal,chen2022marnet}. These works adopt a similar training procedure to that of \inte. Their main novelty comes from the type of the trigger signal being used. However, their trigger types either require the adversary to directly change the agent's observations, which violates the setting of this work, or are designed for specific domains. In comparison, \inte~only declares a general trigger signal in its algorithm. Therefore, it is feasible to extend \inte~to our setting. We also compare with \bdrl~because it also uses the adversary agent's actions to affect others' observations. 

For \inte, it determines whether an episode is abnormal based on: $rand(0,1)>0.5-(P_t-P_c)$ where $rand(0,1)$ is a random value among $(0,1)$, $P_t$ is the attack success rate, and $P_c$ is the task success rate. Moreover, in an abnormal episode, \inte~lets the adversary agent perform the trigger action in a random time step. After the trigger action, \inte~also changes the following rewards to the opposite of their original values.
For \bdrl, it first trains a benign team policy and a malicious team policy. Then, the adversary agent will randomly perform a fixed-length of trigger actions. Before the trigger actions, \bdrl~uses the benign policy to select the teammates' actions while it uses the malicious policy after the trigger actions. Given the collected experience, \bdrl~uses behavior cloning to burn both benign and malicious behaviors into a single policy. 

\subsubsection{Network Architectures and Parameters}
Each teammate $k$'s policy $\pi_k$ is a deep recurrent Q-network. $\pi_k$ and the mixing network have the same architecture as the QMIX implementation in \url{https://github.com/oxwhirl/pymarl}.
The adversary agent policy $\pi_i$ also has the architecture of $\pi_k$ and it has an additional embedding layer that transforms the binary input $d_t\&y$ into a vector whose size is the observation size. This vector is input to the deep recurrent Q-network along with the observation. The trigger policy $\pi^{tri}$ has the same architecture as $\pi_k$ except that its output size is $2$. The trigger value estimator $V^{tri}$ consists of three linear layers with Relu as the activation. Its hidden layer size is 128. The fixed observation encoder $\bar{\mathcal{E}}(\cdot)$ is a single linear layer with the output size as $64$ and the trainable encoder $\mathcal{E}(\cdot)$ consists of three linear layers with Relu as the activation. Its hidden layer size is $128$ and its output size is $64$.

We train all the networks with \emph{RMSprop} optimizer and the learning rate is $0.0005$. The training step limit $T_l= 2e7$ and we update target network every $200$ episodes. The replay buffer size is $5000$, the batch size is $32$, and we run $8$ sampling processes concurrently. The discount factor $\gamma=0.99$. During the training, exploration is based on an $\epsilon$-greedy action selection. We anneal $\epsilon$ linearly from $1$ to $0.05$ over $50000$ steps and keep it constant for the rest of training. We test the trained model every $20000$ steps during the training. We run all experiment for $5$ random seeds in a workstation equipped with an Intel Core i9-10940X CPU and four Nvidia GeForce 2080Ti GPUs.

\begin{figure*}[tp]
	\centering
	\subfigure[Win rate for benign episodes ]{\includegraphics[width=.32\textwidth]{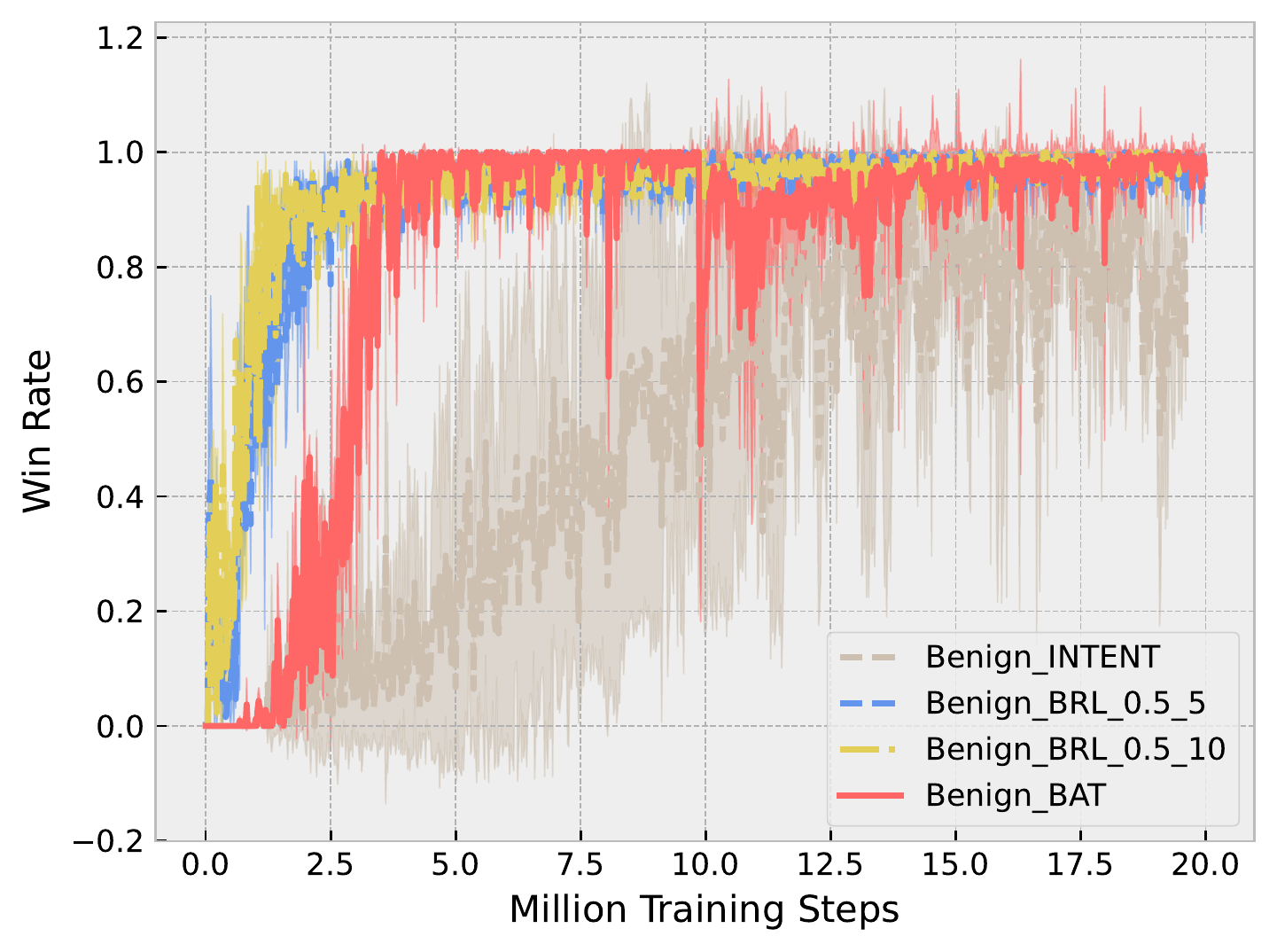}\label{3s3z_benign}}
	\subfigure[Fail rate for abnormal episodes]{\includegraphics[width=.32\textwidth]{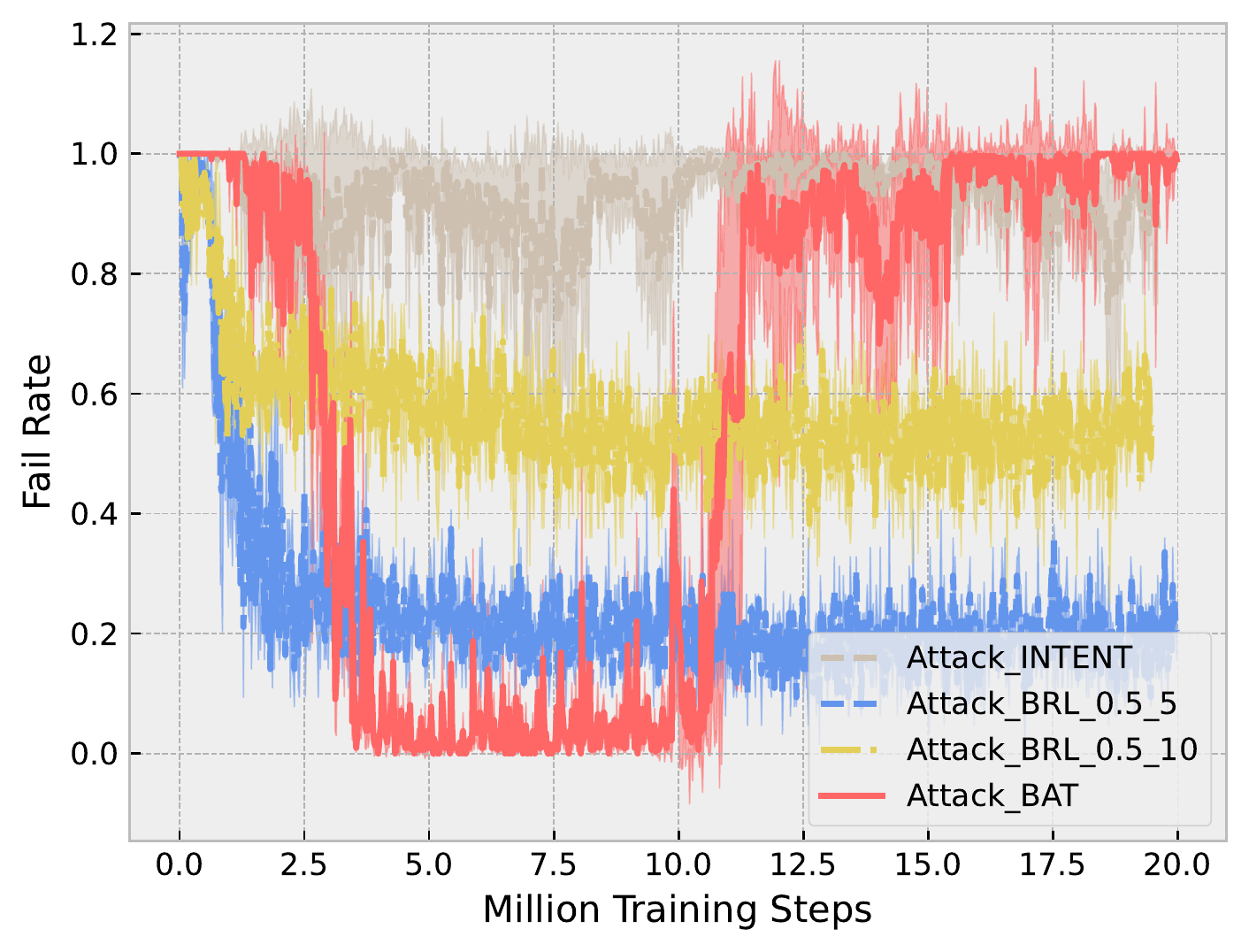}\label{3s3z_attack}}
	\subfigure[Trigger time of each abnormal episode]{\includegraphics[width=.32\textwidth]{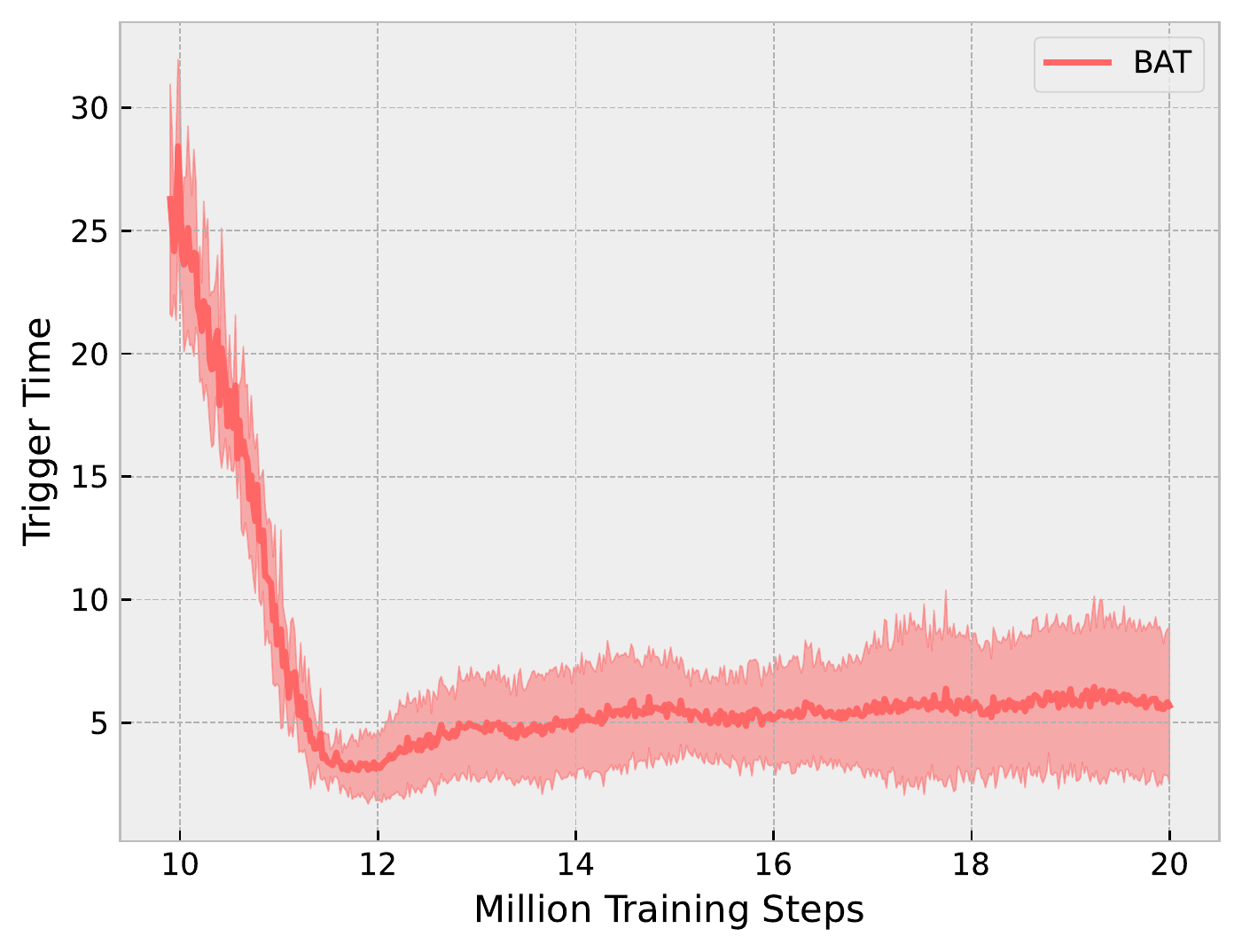}\label{trigger_time}}
	\caption{The training results of the backdoor attack on the  \textit{3s\_vs\_3z} map.}
	\label{fig_com}
\end{figure*}

\subsubsection{Evaluation Results}
To evaluate the performance of the backdoor attack, we mainly consider two metrics, i.e., the team win rate in benign episodes and the team fail rate in abnormal episodes. A successful backdoor attack should achieve both a high benign win rate and a high abnormal fail rate. We compare our attack with \inte~and \bdrl. Because \bdrl~does not specify the percentage of abnormal episodes $P_{tri}$ and the number of trigger actions $N_{tri}$, we set $P_{tri}=0.5$ for \bdrl~which is the same as our setting. We also test two settings, i.e., $N_{tri}=5$ and $N_{tri}=10$ for \bdrl. We denote our attack as $BAT$ (Backdoor Attack on Teamwork), and denote the two settings of \bdrl~as $BRL\_0.5\_5$ and $BRL\_0.5\_10$ respectively. The results are as shown in Figure \ref{fig_com}. Due to the space limit, we only show the results for the \textit{3s\_vs\_3z} map. For the other scenarios, please check the results in Appendix. Figure \ref{3s3z_benign} shows the win rate while Figure \ref{3s3z_attack} shows the fail rate. Note that for the \textit{3s\_vs\_3z} map, the length of our benign training period $T_b=1000000$. Therefore, in the first half of the training, we can see that $BAT$ learns the benign behavior gradually and the fail rate for abnormal episodes in this period is very low. After this benign training period, we start to train the trigger policy and the adversary agent starts to perform trigger actions. From Figure \ref{trigger_time}, we can see that the number of trigger actions performed in each abnormal episode decreases as the training of trigger policy, which means the trigger policy is learning to balance the influence on teammates' observations and the number of trigger times. In the same time, we can observe that our fail rate starts to increase while the win rate does not change a lot. This indicates that the observations of benign episodes are different from those of abnormal episodes and thus, the malicious behavior training does not affect the benign behavior too much.

In comparison, we can observe that the training of \inte~has a very high variance. Note that \inte~adjusts the percentage of abnormal episodes used for training based on the difference between benign performance and malicious performance. The higher the benign performance, the more abnormal episodes will be sampled for training. Figure \ref{fig_com} shows that when \inte~achieves a high win rate, the training with abnormal episodes will degrade its benign performance. This implies that the teammates' observations are not affected much by the trigger action of \inte. Therefore, in abnormal episodes, the teammate policy will output similar Q-value estimations $\bm{Q}_{-i}$ to those in benign episodes due to similar observations. In this case, the reward manipulation of \inte~for abnormal episodes encourages to learn low $\bm{Q}_{-i}$ while the training with benign episodes encourages to learn high $\bm{Q}_{-i}$ for similar observations. This means the training of benign behaviors and malicious behaviors interfere with each other, which damages the overall training.
When it comes to \bdrl, we find that its win rate is always very high while it cannot achieve a high fail rate for abnormal episodes. We think this also results from the fact that the trigger actions of \bdrl~do not affect the teammates' observations efficiently. Because the observations of benign and abnormal episodes are still quite similar, the behavior cloning for benign behaviors can dominate the behavior cloning for abnormal behaviors and thus, prevent the fail rate from increasing. Actually, this hypothesis can be supported by Figure \ref{3s3z_attack}. When \bdrl~performs $10$ trigger actions instead of just $5$, it can obtain a higher fail rate. This is because $10$ trigger actions can affect the teammates' observations to a larger extent and thus, reduce the conflicts between the behavior cloning for benign and malicious behaviors.

\subsubsection{Ablation Study} 
\begin{figure}[t]
	\centering
	\includegraphics[width=0.8\columnwidth]{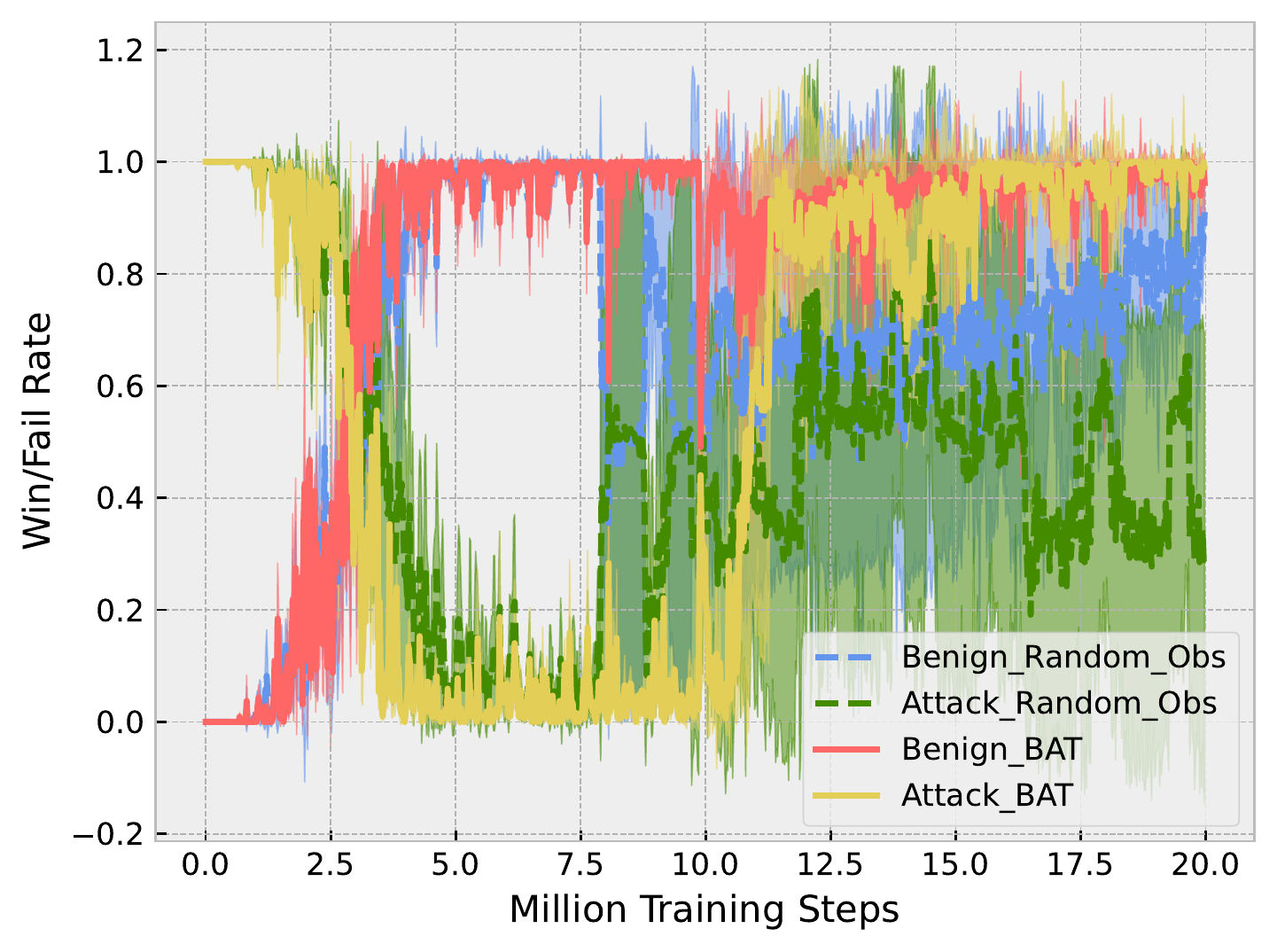} 
	\caption{The ablation study for $r_{obs}$.}
	\label{fig_ablation}
\end{figure}

In this part, we conduct an ablation study to evaluate the effects of $r_{obs}$.
Specifically, we simply set the $\{r_{obs}\}$ as random values that belongs to $(0,1)$. By comparing this setting with the original setting, we can investigate whether the $\{r_{obs}\}$ computed by the RND module really provide meaningful information for the training. We denote the ablation study setting as \emph{Rondom\_Obs} and present the results in Figure \ref{fig_ablation}. We show both the win rate performance and fail rate performance in the same figure. From this figure, we can clearly see that the random $r_{obs}$ totally destroys the backdoor training. The resulting team policy shows high-variance performances for both benign and abnormal episodes. The reason is that the random $r_{obs}$ cannot provide useful information for training the trigger policy. This means the trigger policy may output random trigger decisions, which increases the conflicts between benign behavior training and malicious behavior training. Therefore, this study demonstrates that the $\{r_{obs}\}$ is an essential part for our methods. Based on the information provided by $\{r_{obs}\}$, the trigger policy can learn to efficiently affect the teammates' observations with only a few actions. 

\section{Conclusions}
In this work, we propose a backdoor attack on multiagent collaborative systems. To make the attack more practical, we aim to let the adversary agent use its actions to trigger the backdoor of teammates' policy. Since these systems are often partially observable, whether an action can affect the teammates are uncertain. We identify the issue that when the adversary agent cannot affect its teammates' observations efficiently, the similar observations of benign and abnormal episodes can damage the backdoor training. To solve this issue, we propose to learn the auxiliary rewards to measure the influences on teammates' observations. Based on these rewards, we train a trigger policy that instructs the adversary agent to perform meaningful trigger actions. This further facilitates the backdoor training. We have conducted extensive experiments to demonstrate the effectiveness of our attack. The results show that the trained team policy can complete the task well when without the trigger actions while a few trigger actions performed by the adversary agent can degrade the teamwork performance significantly.

\bibliography{aaai23}

\begin{thebibliography}{23}
\providecommand{\natexlab}[1]{#1}

\bibitem[{Ashcraft and Karra(2021)}]{ashcraftPoisoning}
Ashcraft, C.; and Karra, K. 2021.
\newblock Poisoning Deep Reinforcement Learning Agents with In-Distribution
  Triggers.
\newblock arXiv:2106.07798.

\bibitem[{Bhagoji et~al.(2019)Bhagoji, Chakraborty, Mittal, and
  Calo}]{bhagoji2019analyzing}
Bhagoji, A.~N.; Chakraborty, S.; Mittal, P.; and Calo, S. 2019.
\newblock Analyzing federated learning through an adversarial lens.
\newblock In \emph{International Conference on Machine Learning}, 634--643.
  PMLR.

\bibitem[{Burda et~al.(2019)Burda, Edwards, Storkey, and
  Klimov}]{burda2018exploration}
Burda, Y.; Edwards, H.; Storkey, A.; and Klimov, O. 2019.
\newblock Exploration by random network distillation.
\newblock In \emph{International Conference on Learning Representations}.

\bibitem[{Chen et~al.(2017)Chen, Liu, Li, Lu, and Song}]{chen2017targeted}
Chen, X.; Liu, C.; Li, B.; Lu, K.; and Song, D. 2017.
\newblock Targeted backdoor attacks on deep learning systems using data
  poisoning.
\newblock arXiv:1712.05526.

\bibitem[{Chen et~al.(2021)Chen, Salem, Backes, Ma, and Zhang}]{chen2021badnl}
Chen, X.; Salem, A.; Backes, M.; Ma, S.; and Zhang, Y. 2021.
\newblock Badnl: Backdoor attacks against nlp models.
\newblock In \emph{ICML 2021 Workshop on Adversarial Machine Learning}.

\bibitem[{Chen, Zheng, and Gong(2022)}]{chen2022marnet}
Chen, Y.; Zheng, Z.; and Gong, X. 2022.
\newblock MARNet: Backdoor Attacks Against Cooperative Multi-Agent
  Reinforcement Learning.
\newblock \emph{IEEE Transactions on Dependable and Secure Computing}.

\bibitem[{Gu, Dolan-Gavitt, and Garg(2017)}]{gu2017badnets}
Gu, T.; Dolan-Gavitt, B.; and Garg, S. 2017.
\newblock BadNets: Identifying Vulnerabilities in the Machine Learning Model
  Supply Chain.
\newblock arXiv:1708.06733.

\bibitem[{Hernandez-Leal, Kartal, and Taylor(2019)}]{hernandez2019survey}
Hernandez-Leal, P.; Kartal, B.; and Taylor, M.~E. 2019.
\newblock A survey and critique of multiagent deep reinforcement learning.
\newblock \emph{Autonomous Agents and Multi-Agent Systems}, 33(6): 750--797.

\bibitem[{Kiourti et~al.(2019)Kiourti, Wardega, Jha, and Li}]{kiourtiTrojDRL}
Kiourti, P.; Wardega, K.; Jha, S.; and Li, W. 2019.
\newblock TrojDRL: Trojan Attacks on Deep Reinforcement Learning Agents.

\bibitem[{Liu et~al.(2017)Liu, Ma, Aafer, Lee, Zhai, Wang, and
  Zhang}]{liu2017trojaning}
Liu, Y.; Ma, S.; Aafer, Y.; Lee, W.-C.; Zhai, J.; Wang, W.; and Zhang, X. 2017.
\newblock Trojaning attack on neural networks.
\newblock Technical report.

\bibitem[{Mnih et~al.(2015)Mnih, Kavukcuoglu, Silver, Rusu, Veness, Bellemare,
  Graves, Riedmiller, Fidjeland, Ostrovski et~al.}]{mnih2015human}
Mnih, V.; Kavukcuoglu, K.; Silver, D.; Rusu, A.~A.; Veness, J.; Bellemare,
  M.~G.; Graves, A.; Riedmiller, M.; Fidjeland, A.~K.; Ostrovski, G.; et~al.
  2015.
\newblock Human-level control through deep reinforcement learning.
\newblock \emph{nature}, 518(7540): 529--533.

\bibitem[{Oliehoek and Amato(2016)}]{oliehoek2016concise}
Oliehoek, F.~A.; and Amato, C. 2016.
\newblock \emph{A concise introduction to decentralized POMDPs}.
\newblock Springer.

\bibitem[{Oliehoek, Spaan, and Vlassis(2008)}]{oliehoek2008optimal}
Oliehoek, F.~A.; Spaan, M.~T.; and Vlassis, N. 2008.
\newblock Optimal and approximate Q-value functions for decentralized POMDPs.
\newblock \emph{Journal of Artificial Intelligence Research}, 32: 289--353.

\bibitem[{Rashid et~al.(2018)Rashid, Samvelyan, Schroeder, Farquhar, Foerster,
  and Whiteson}]{qmix}
Rashid, T.; Samvelyan, M.; Schroeder, C.; Farquhar, G.; Foerster, J.; and
  Whiteson, S. 2018.
\newblock {QMIX}: Monotonic Value Function Factorisation for Deep Multi-Agent
  Reinforcement Learning.
\newblock In Dy, J.; and Krause, A., eds., \emph{Proceedings of the 35th
  International Conference on Machine Learning}, volume~80 of \emph{Proceedings
  of Machine Learning Research}, 4295--4304. PMLR.

\bibitem[{Salem et~al.(2020)Salem, Sautter, Backes, Humbert, and
  Zhang}]{salem2020baaan}
Salem, A.; Sautter, Y.; Backes, M.; Humbert, M.; and Zhang, Y. 2020.
\newblock Baaan: Backdoor attacks against autoencoder and gan-based machine
  learning models.
\newblock arXiv:2010.03007.

\bibitem[{Samvelyan et~al.(2019)Samvelyan, Rashid, Schroeder~de Witt, Farquhar,
  Nardelli, Rudner, Hung, Torr, Foerster, and Whiteson}]{samvelyanSMAC}
Samvelyan, M.; Rashid, T.; Schroeder~de Witt, C.; Farquhar, G.; Nardelli, N.;
  Rudner, T. G.~J.; Hung, C.-M.; Torr, P. H.~S.; Foerster, J.; and Whiteson, S.
  2019.
\newblock The StarCraft Multi-Agent Challenge.
\newblock In \emph{Proceedings of the 18th International Conference on
  Autonomous Agents and MultiAgent Systems}, AAMAS '19, 2186–2188.

\bibitem[{Sutton and Barto(2018)}]{sutton2018reinforcement}
Sutton, R.~S.; and Barto, A.~G. 2018.
\newblock \emph{Reinforcement learning: An introduction}.
\newblock MIT press.

\bibitem[{Wang et~al.(2020)Wang, Sreenivasan, Rajput, Vishwakarma, Agarwal,
  Sohn, Lee, and Papailiopoulos}]{wang2020attack}
Wang, H.; Sreenivasan, K.; Rajput, S.; Vishwakarma, H.; Agarwal, S.; Sohn,
  J.-y.; Lee, K.; and Papailiopoulos, D. 2020.
\newblock Attack of the tails: Yes, you really can backdoor federated learning.
\newblock \emph{Advances in Neural Information Processing Systems}, 33:
  16070--16084.

\bibitem[{Wang et~al.(2021{\natexlab{a}})Wang, Javed, Wu, Guo, Xing, and
  Song}]{wangBACKDOORL}
Wang, L.; Javed, Z.; Wu, X.; Guo, W.; Xing, X.; and Song, D.
  2021{\natexlab{a}}.
\newblock BACKDOORL: Backdoor Attack against Competitive Reinforcement
  Learning.
\newblock In Zhou, Z.-H., ed., \emph{Proceedings of the Thirtieth International
  Joint Conference on Artificial Intelligence, {IJCAI-21}}, 3699--3705.
  International Joint Conferences on Artificial Intelligence Organization.
\newblock Main Track.

\bibitem[{Wang et~al.(2021{\natexlab{b}})Wang, Sarkar, Li, Maniatakos, and
  Jabari}]{wangTransport}
Wang, Y.; Sarkar, E.; Li, W.; Maniatakos, M.; and Jabari, S.~E.
  2021{\natexlab{b}}.
\newblock Stop-and-Go: Exploring Backdoor Attacks on Deep Reinforcement
  Learning-Based Traffic Congestion Control Systems.
\newblock \emph{Trans. Info. For. Sec.}, 16: 4772–4787.

\bibitem[{Xie et~al.(2019)Xie, Huang, Chen, and Li}]{xie2019dba}
Xie, C.; Huang, K.; Chen, P.-Y.; and Li, B. 2019.
\newblock Dba: Distributed backdoor attacks against federated learning.
\newblock In \emph{International Conference on Learning Representations}.

\bibitem[{Yang et~al.(2019)Yang, Iyer, Reimann, and Virani}]{yangIntentional}
Yang, Z.; Iyer, N.; Reimann, J.; and Virani, N. 2019.
\newblock Design of intentional backdoors in sequential models.
\newblock arXiv:1902.09972.

\bibitem[{Yu et~al.(2022)Yu, Liu, Li, Huang, and Feng}]{yu2022temporal}
Yu, Y.; Liu, J.; Li, S.; Huang, K.; and Feng, X. 2022.
\newblock A Temporal-Pattern Backdoor Attack to Deep Reinforcement Learning.
\newblock arXiv:2205.02589.

\end{thebibliography}

\appendix


\newpage
\onecolumn

\section{Appendix}
In the appendix, we first show the comparison results for the remaining scenarios, i.e., \textit{8m}, \textit{MMM0}, \textit{MMM3}, and \textit{MMM9}. Then, we show the experiment results that indicate the influences of the adversary agent on its teammates. Note that we have set different lengths of the benign training period $T_b$ for difference scenarios. For \textit{3s\_vs\_3z}, $T_b=1000000$ while $T_b=4000000$ for \textit{8m}, $T_b=8000000$ for \textit{MMM0}, $T_b=8000000$ for \textit{MMM3}, and $T_b=6000000$ for \textit{MMM9}.

\subsection{Comparison Results for the Remaining Scenarios}
\begin{figure*}[!b]
	\centering
	\subfigure[Win rate for benign episodes ]{\includegraphics[width=.32\textwidth, height=4.25cm]{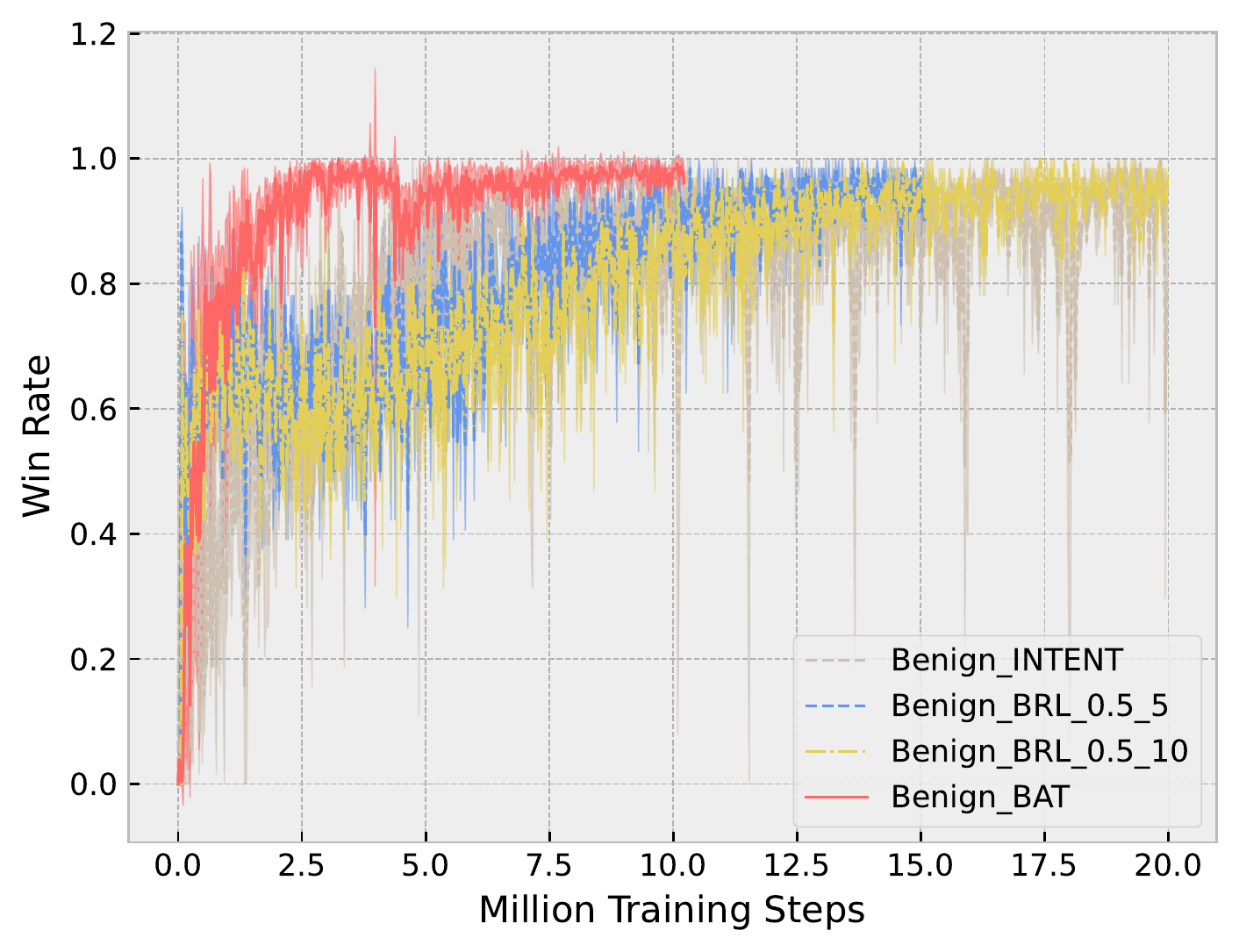}\label{8m_benign}}
	\subfigure[Fail rate for abnormal episodes]{\includegraphics[width=.32\textwidth, height=4.25cm]{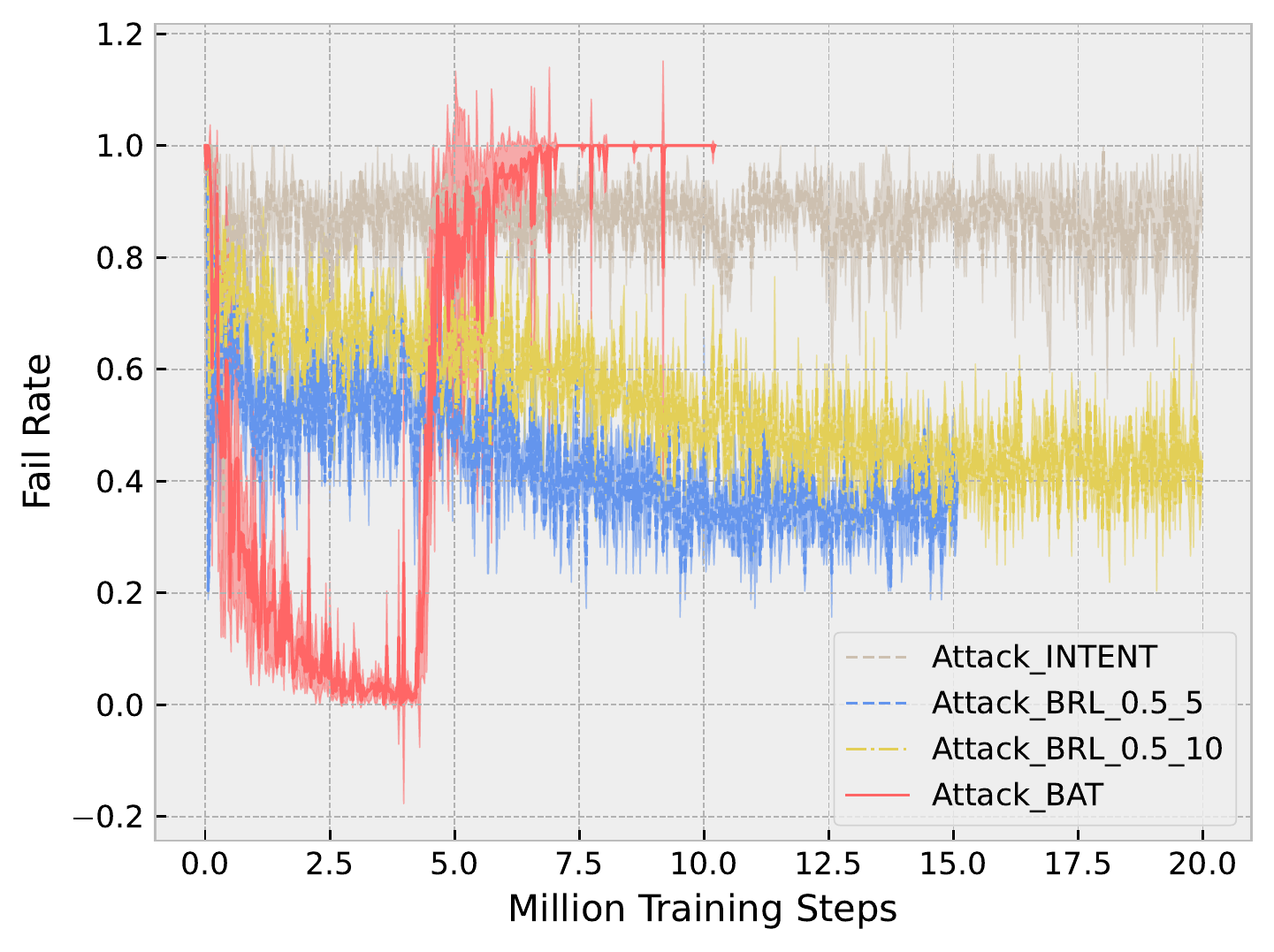}\label{8m_attack}}
	\subfigure[Trigger time of each abnormal episode]{\includegraphics[width=.32\textwidth, height=4.25cm]{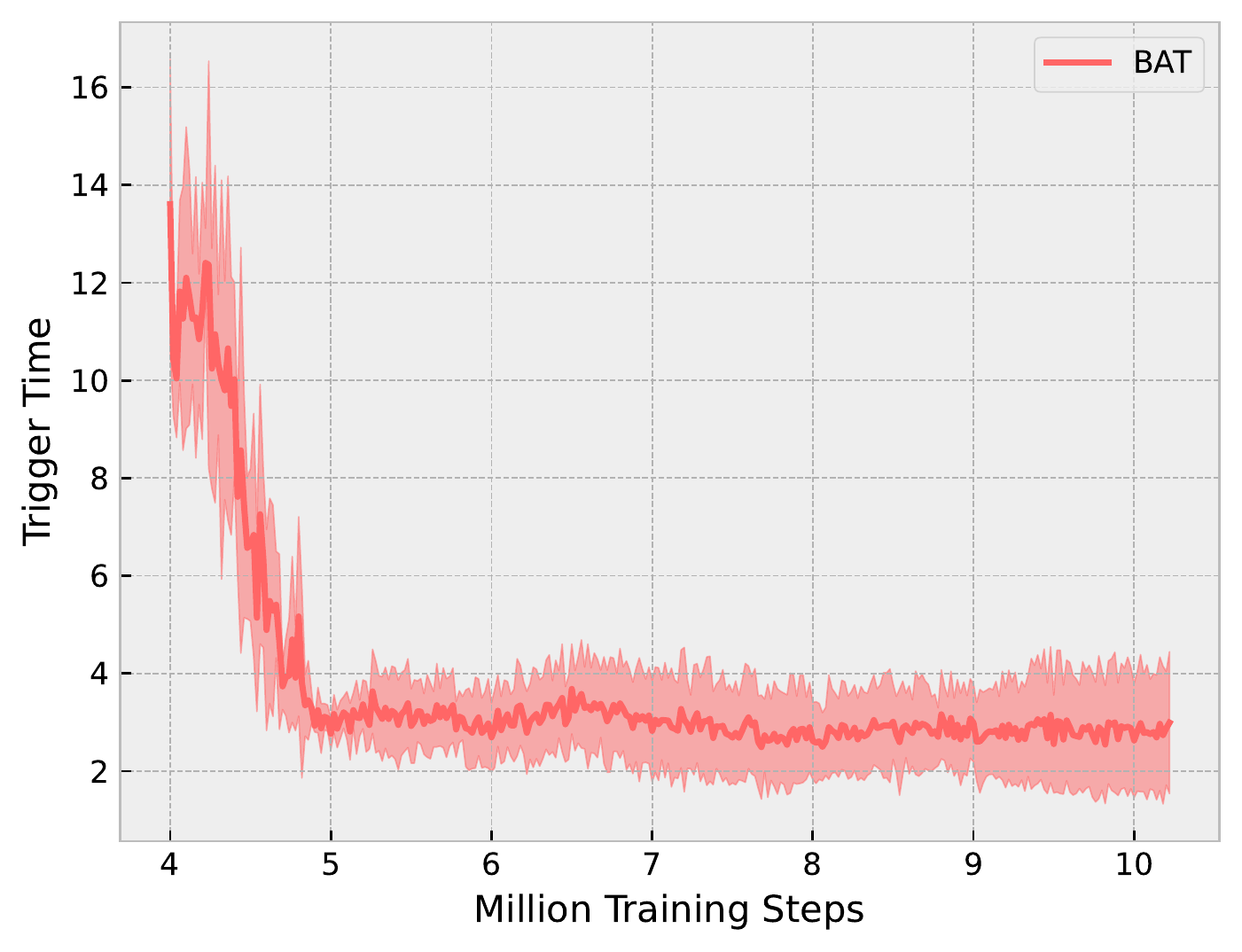}\label{8m_trigger_time}}
	\caption{The training results of the backdoor attack on the  \textit{8m} scenario.}
	\label{fig_8m}
\end{figure*}

\begin{figure*}[!b]
	\centering
	\subfigure[Win rate for benign episodes ]{\includegraphics[width=.32\textwidth, height=4.25cm]{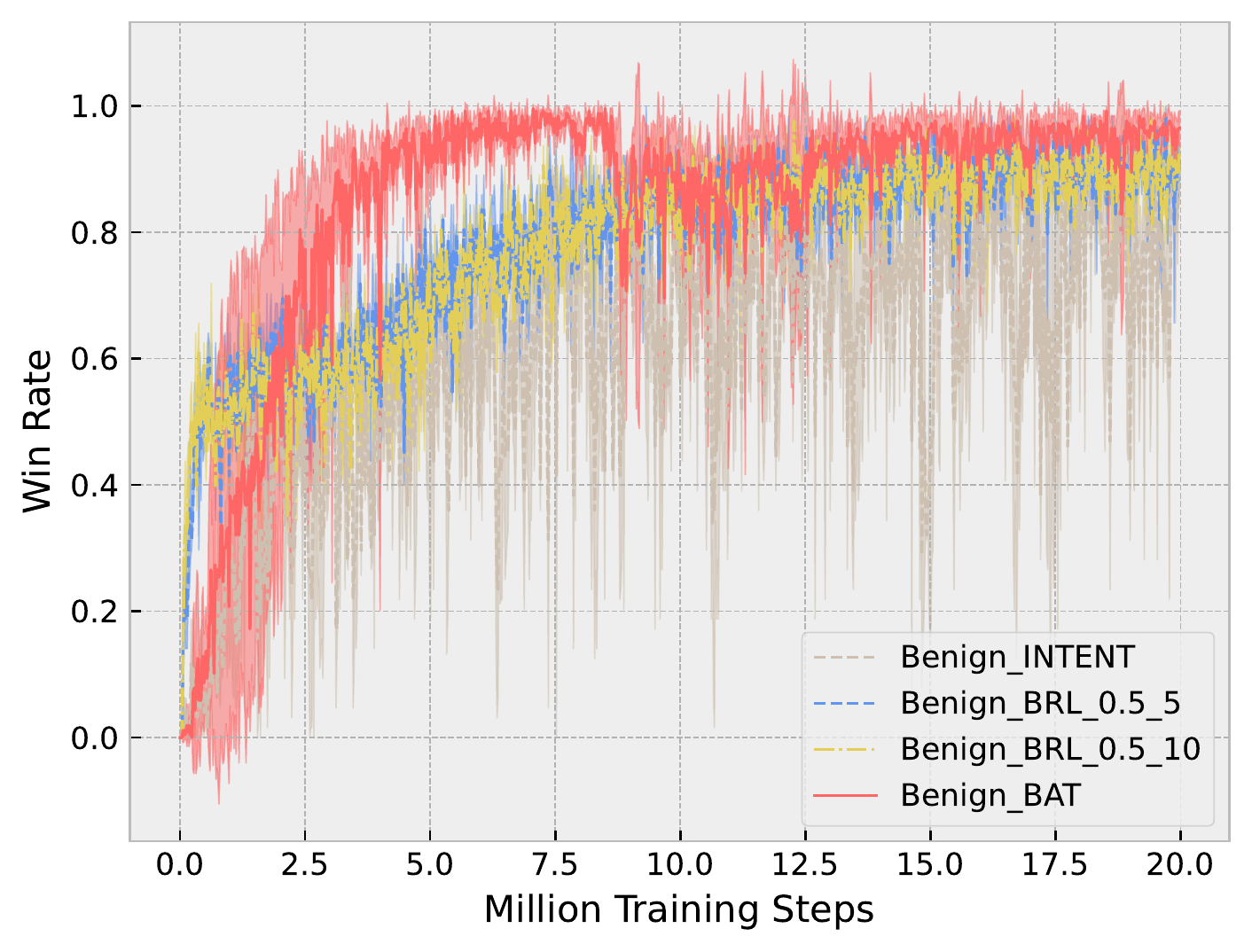}\label{mmm0_benign}}
	\subfigure[Fail rate for abnormal episodes]{\includegraphics[width=.32\textwidth, height=4.25cm]{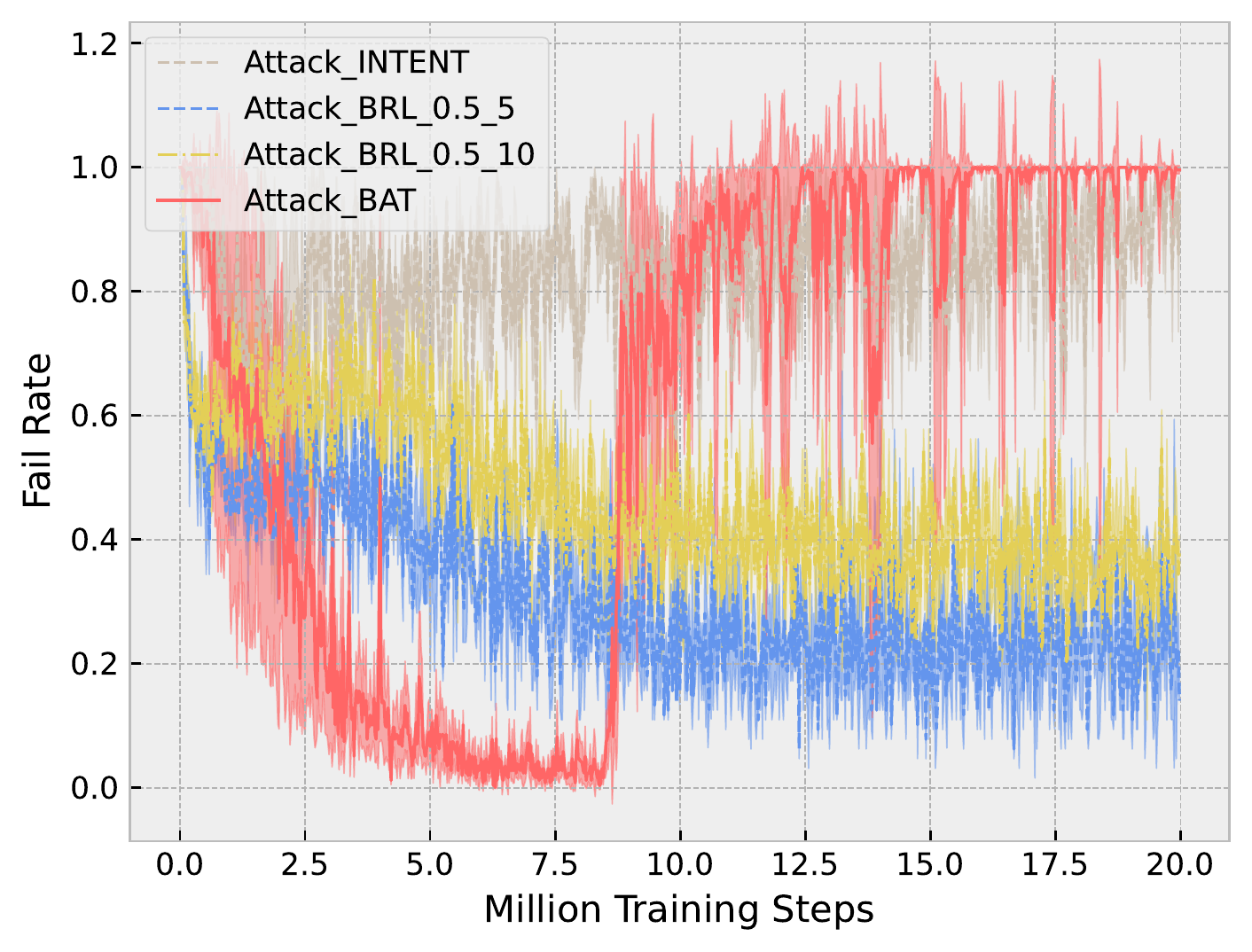}\label{mmm0_attack}}
	\subfigure[Trigger time of each abnormal episode]{\includegraphics[width=.32\textwidth, height=4.25cm]{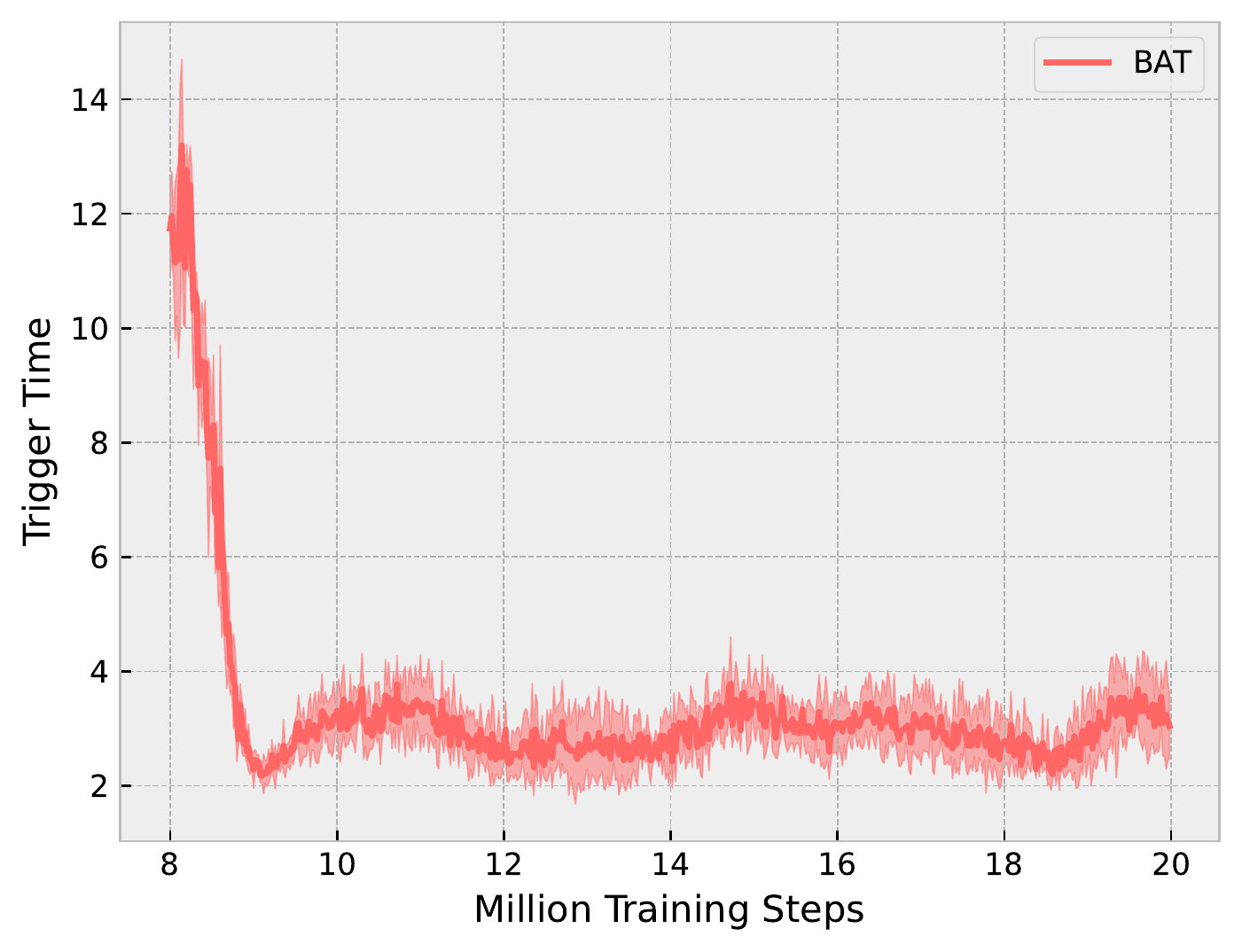}\label{mmm0_trigger_time}}
	\caption{The training results of the backdoor attack on the  \textit{MMM0} scenario.}
	\label{fig_mmm0}
\end{figure*}

\begin{figure*}[!b]
	\centering
	\subfigure[Win rate for benign episodes ]{\includegraphics[width=.32\textwidth, height=4.25cm]{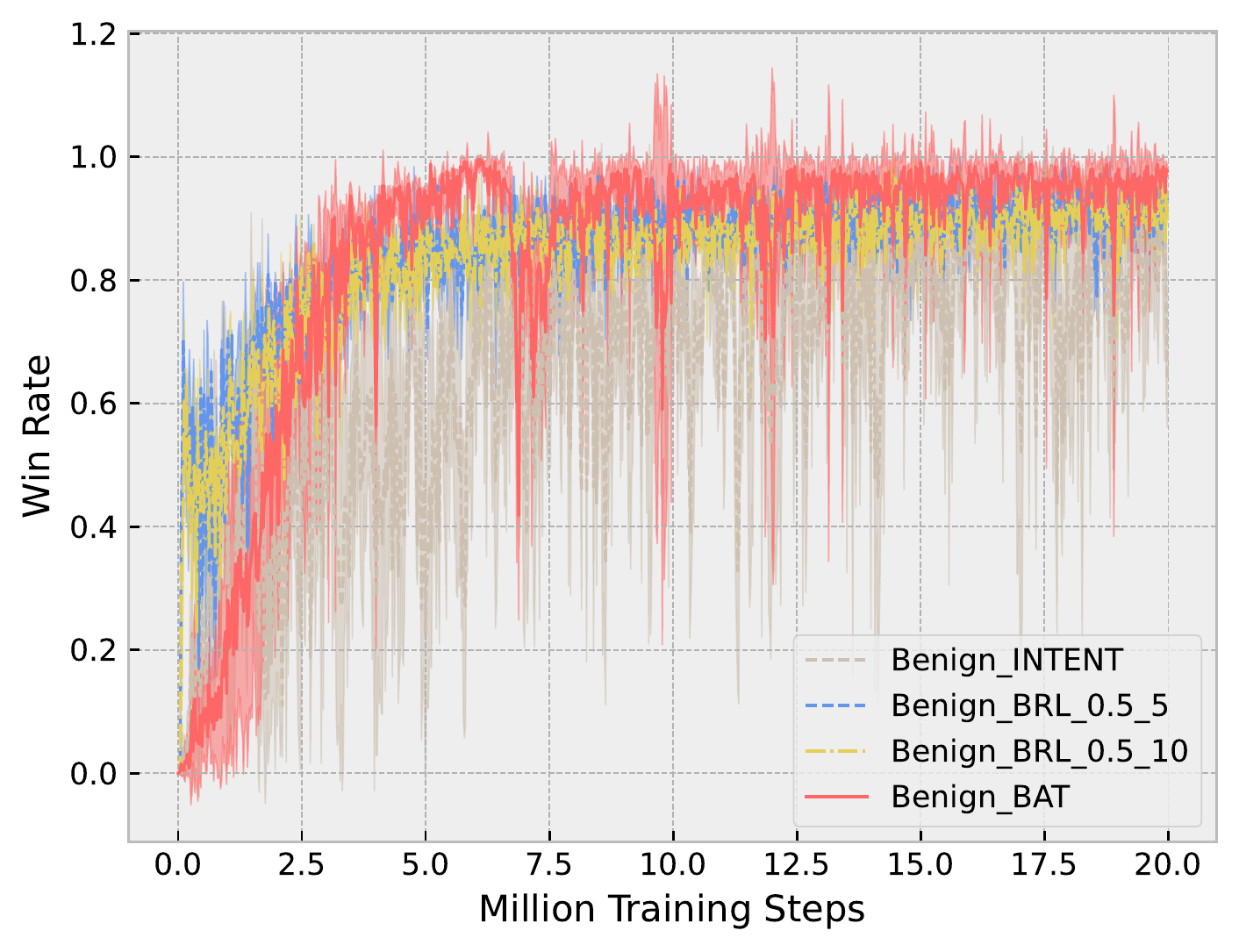}\label{mmm9_benign}}
	\subfigure[Fail rate for abnormal episodes]{\includegraphics[width=.32\textwidth, height=4.25cm]{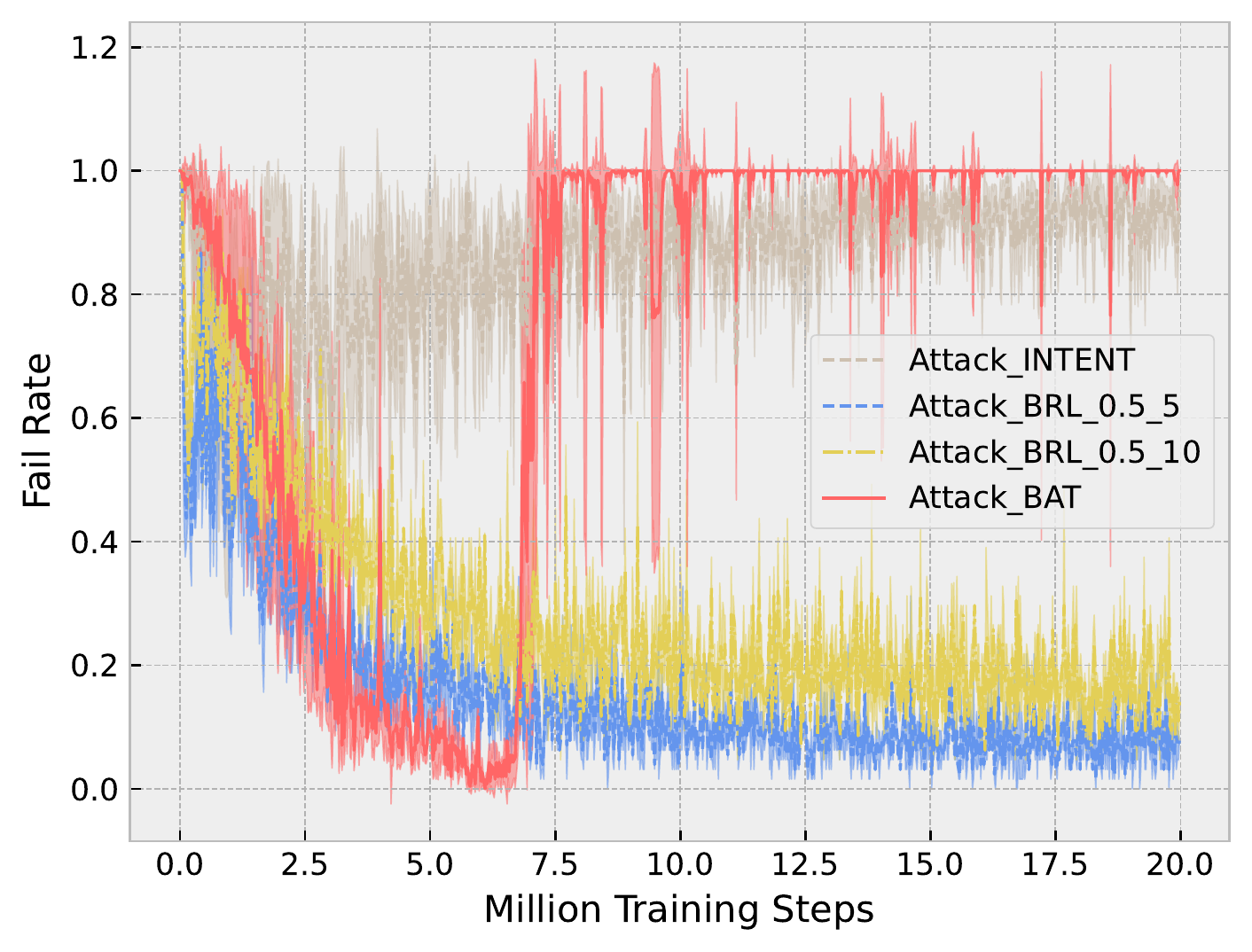}\label{mmm9_attack}}
	\subfigure[Trigger time of each abnormal episode]{\includegraphics[width=.32\textwidth, height=4.25cm]{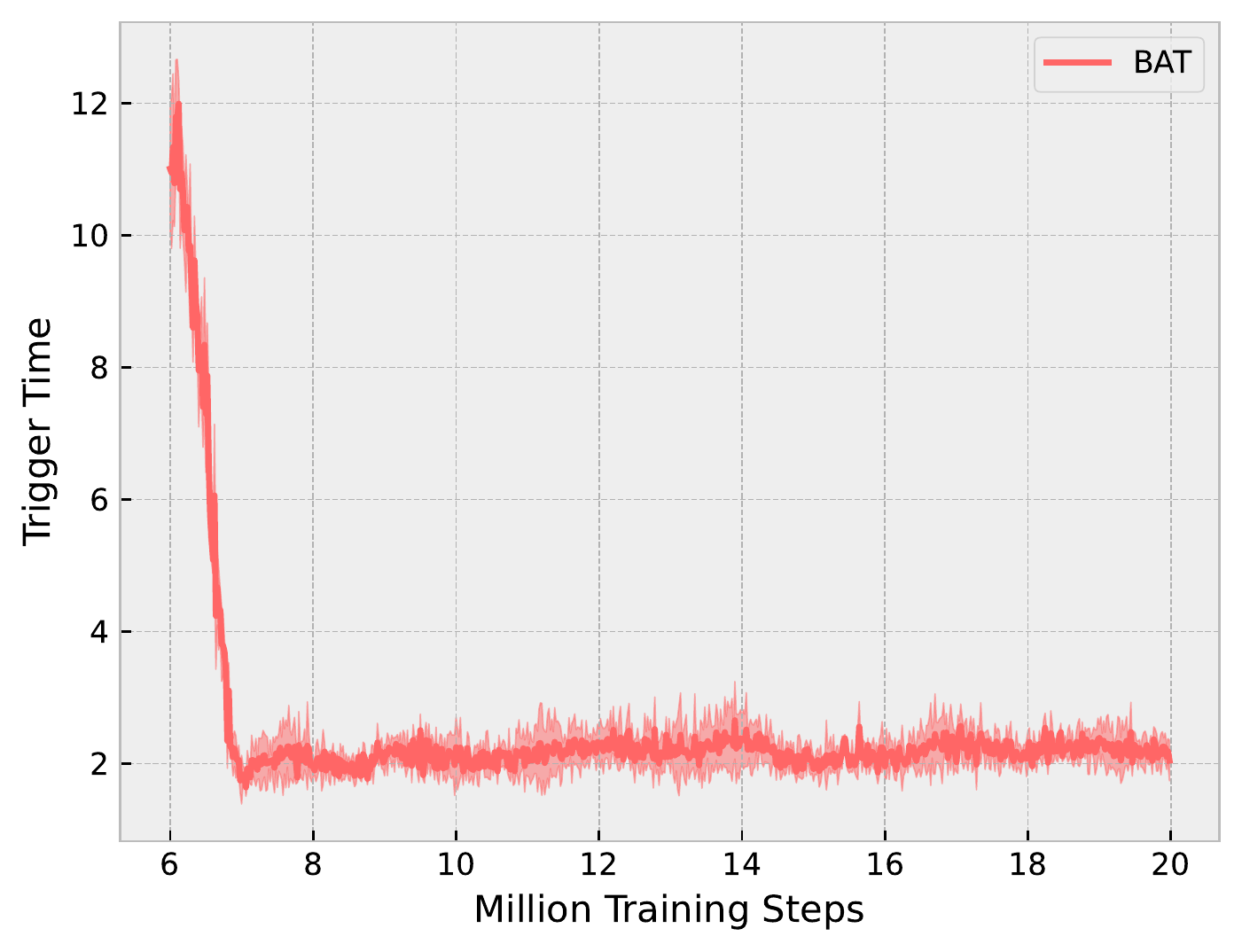}\label{mmm9_trigger_time}}
	\caption{The training results of the backdoor attack on the  \textit{MMM9} scenario.}
	\label{fig_mmm9}
\end{figure*}

\begin{figure*}[tp]
	\centering
	\subfigure[Win rate for benign episodes ]{\includegraphics[width=.32\textwidth, height=4.25cm]{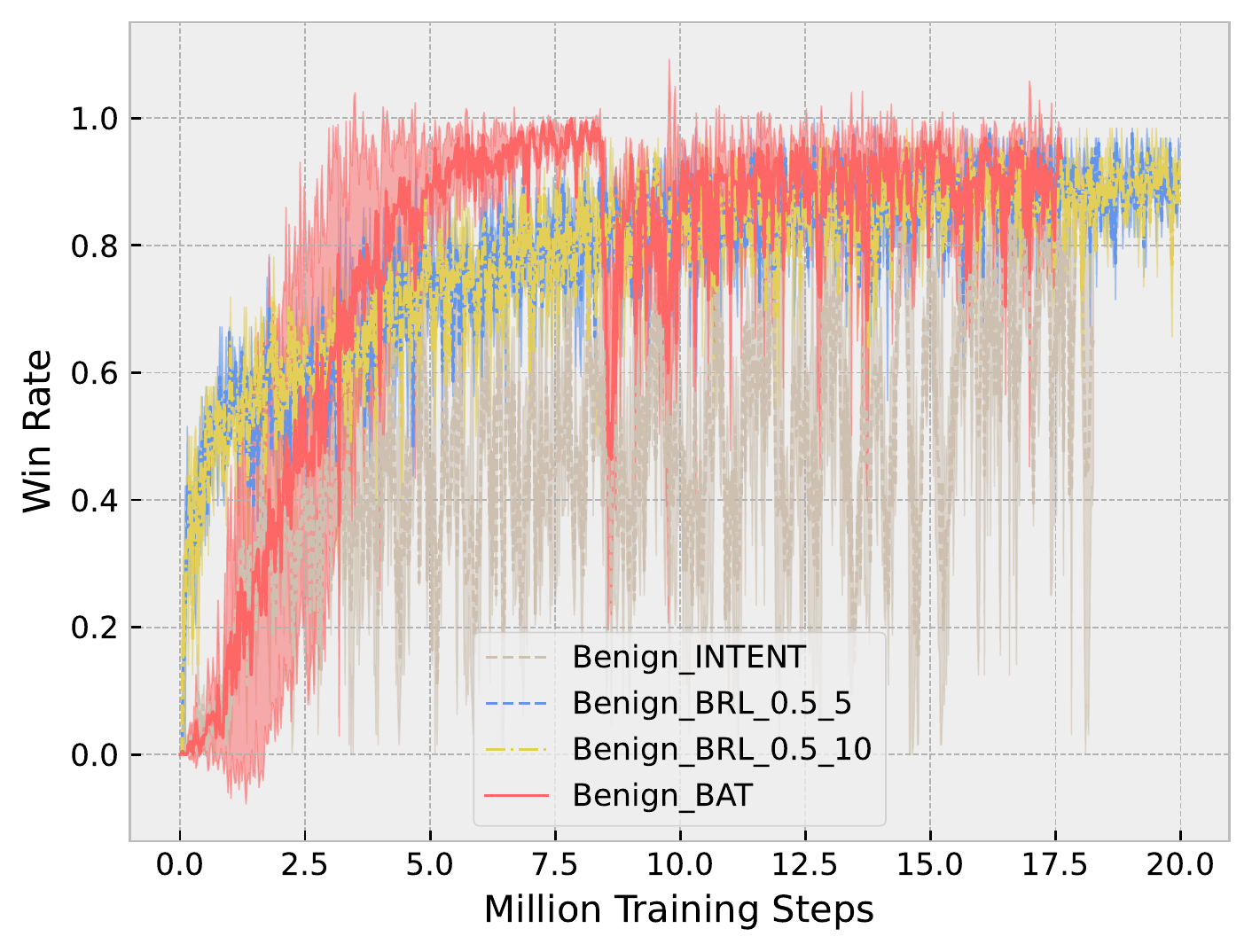}\label{mmm3_benign}}
	\subfigure[Fail rate for abnormal episodes]{\includegraphics[width=.32\textwidth, height=4.25cm]{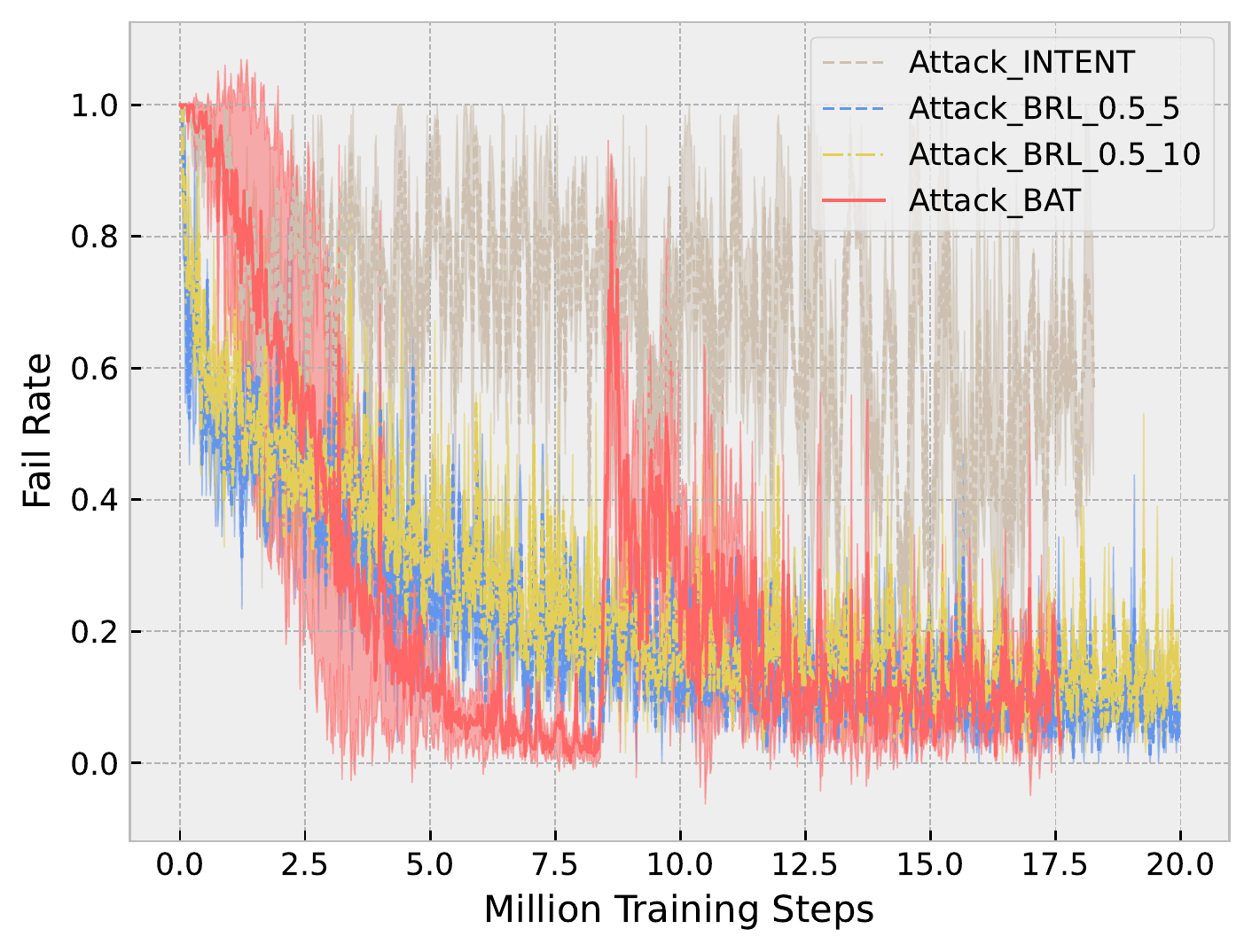}\label{mmm3_attack}}
	\subfigure[Trigger time of each abnormal episode]{\includegraphics[width=.32\textwidth, height=4.25cm]{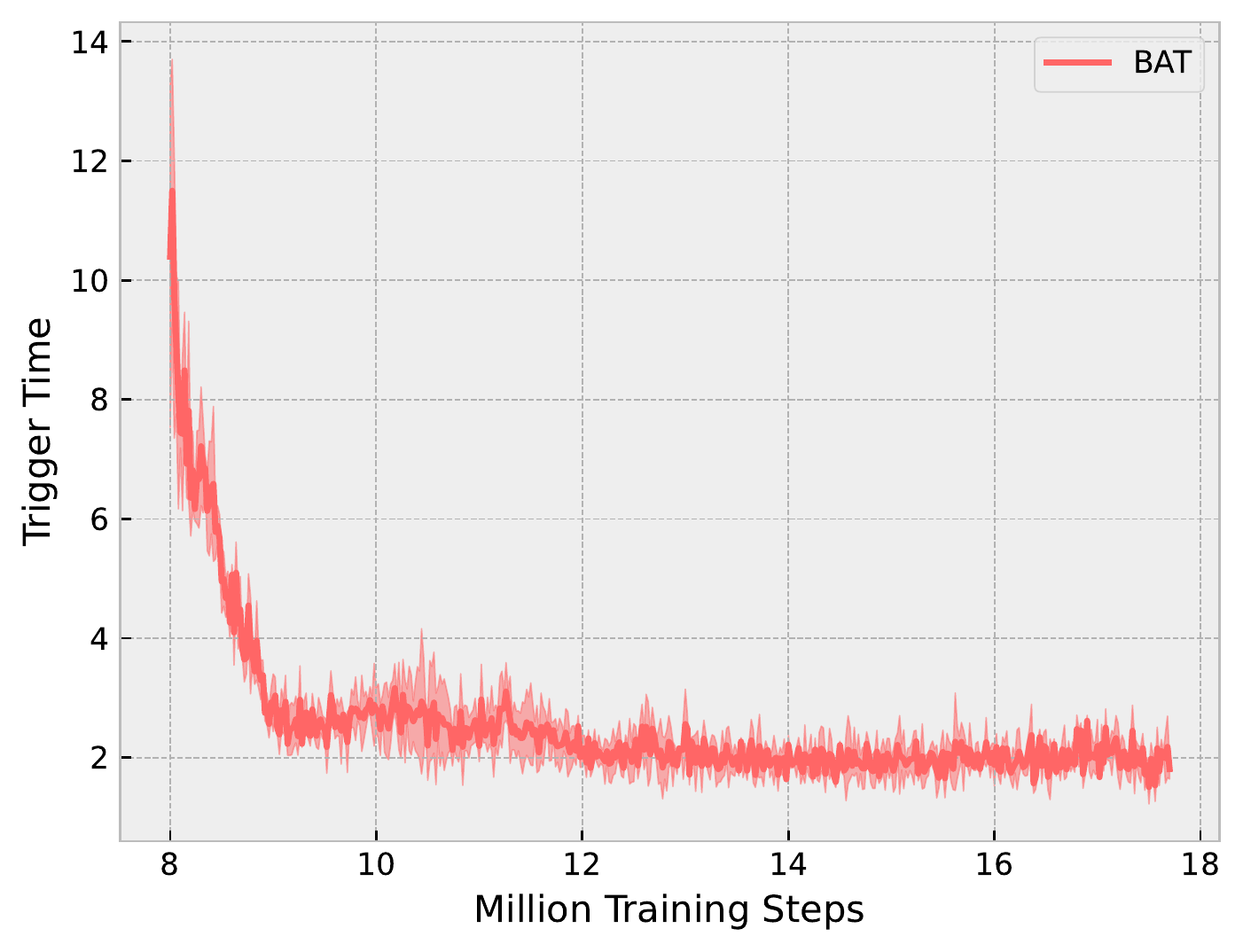}\label{mmm3_trigger_time}}
	\caption{The training results of the backdoor attack on the  \textit{MMM3} scenario.}
	\label{fig_mmm3}
\end{figure*}

The comparison results are shown as follows. Note that sometimes we terminate the training before $T_l= 2e7$ steps if the training has converged. From Figure \ref{fig_8m}, \ref{fig_mmm0}, and \ref{fig_mmm9}, we can see BAT and the baseline methods all achieve similar results to those in the \textit{3s\_vs\_3z} map. This demonstrates the effectiveness of BAT in different scenarios.
However, we also find that BAT only achieves comparable performance to that of \bdrl~in the \textit{MMM3} scenario as shown in Figure \ref{fig_mmm3}. This is because in the \textit{MMM3} scenario, the adversary agent controls a marine, which cannot place enough influences on its teammates' observations. Note that in the \textit{MMM} map, the victim team consists of seven
marines, two marauders, and one medivac. In the \textit{MMM0} scenario, the adversary agent controls one marauder while in the \textit{MMM9} scenario, the adversary agent controls the medivac. Both marauder and medivac play a more important role than the marine in the team. Therefore, the adversary agent in the \textit{MMM3} scenario may not be able to affect its teammates' observations as greatly as in the other two scenarios. 

This also can be identified based on the performance of \inte~and \bdrl. From Figure \ref{fig_mmm3}, we can see that \inte~in the \textit{MMM3} scenario shows a higher variance for both win rate and fail rate than in the other scenarios. This is because the observation difference is smaller and the interference between the benign behavior training and malicious behavior training is larger. Also, \bdrl~generally can improve its fail rate by increasing the number of trigger actions. However, in Figure \ref{mmm3_attack}, we can find that the improvement of fail rate from $BRL\_0.5\_5$ to $BRL\_0.5\_10$ is much smaller than those in the other scenarios. This also indicates that the marine has a weaker ability to affect its teammates and thus, the benefit of performing more trigger actions becomes smaller.
Therefore, the results in Figure \ref{fig_mmm3} demonstrate that the role of the adversary agent in a heterogeneous team can affect the training of backdoor attacks. To launch a backdoor attack easily, the adversary should choose to control the type of agent that has a stronger ability to affect its teammates. Furthermore, to launch the backdoor attack in the \textit{MMM3} scenario, we may need to fine tune the hyper-parameters or revise the method to enable the adversary agent to affect its teammates more effectively. 

\subsection{Results about the Influences on Teammates}
\begin{figure*}[!t]
	\centering
	\subfigure[\textit{3s\_vs\_3z} ]{\includegraphics[width=.32\textwidth, height=4.25cm]{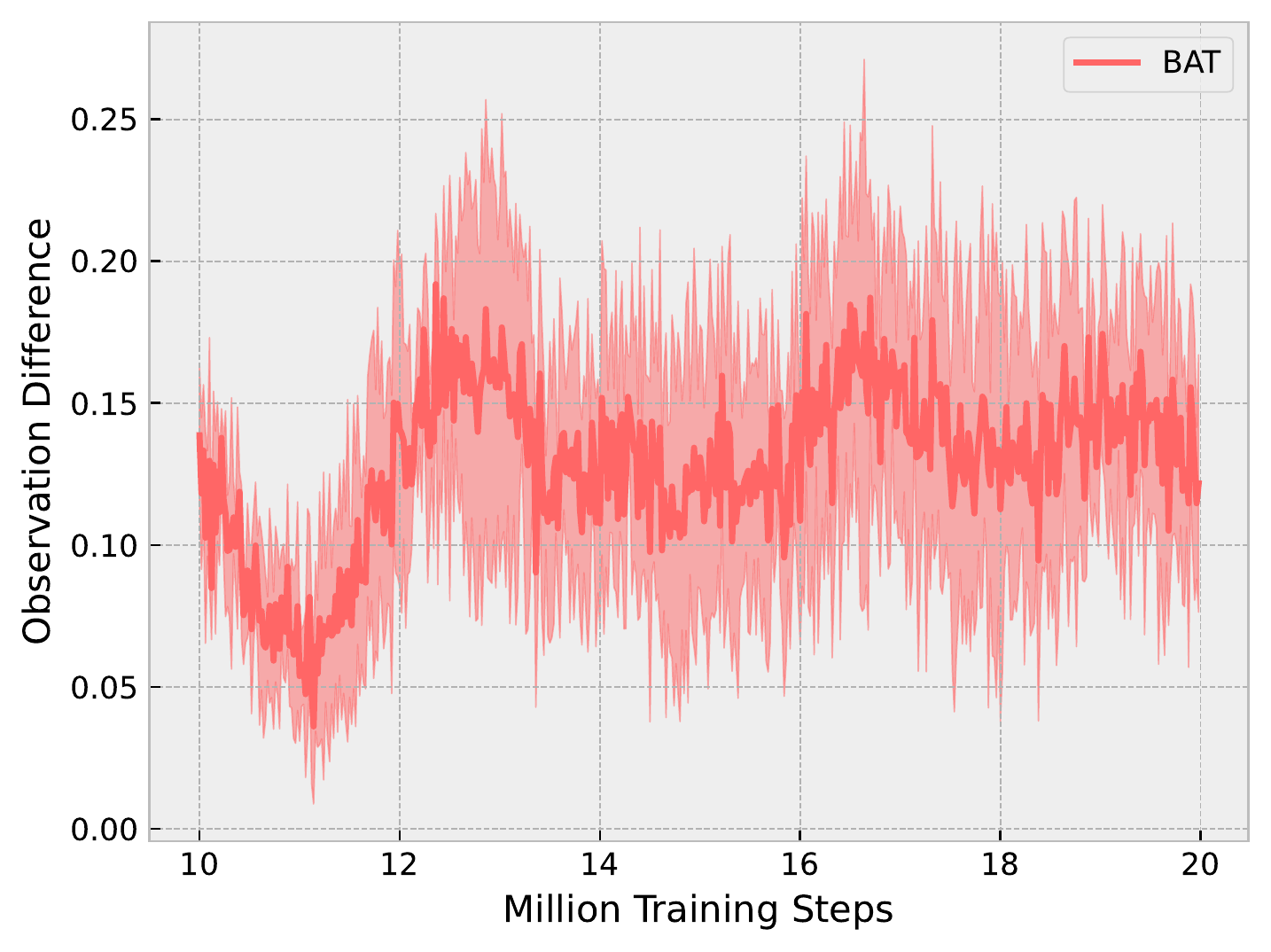}\label{3s3z_obs_diff}}
	\subfigure[\textit{8m}]{\includegraphics[width=.32\textwidth, height=4.25cm]{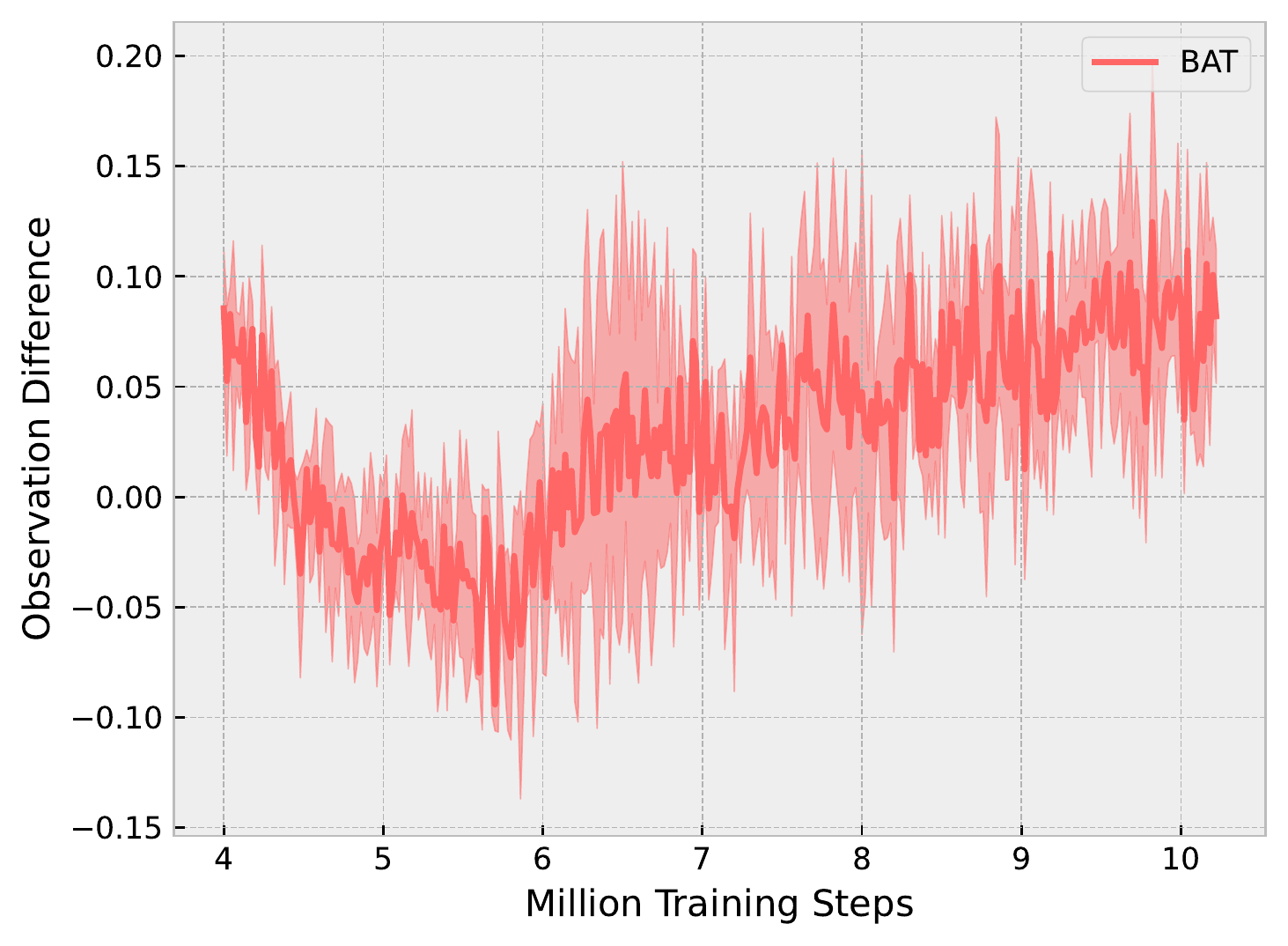}\label{8m_obs_diff}}
	\subfigure[\textit{MMM0}]{\includegraphics[width=.32\textwidth, height=4.25cm]{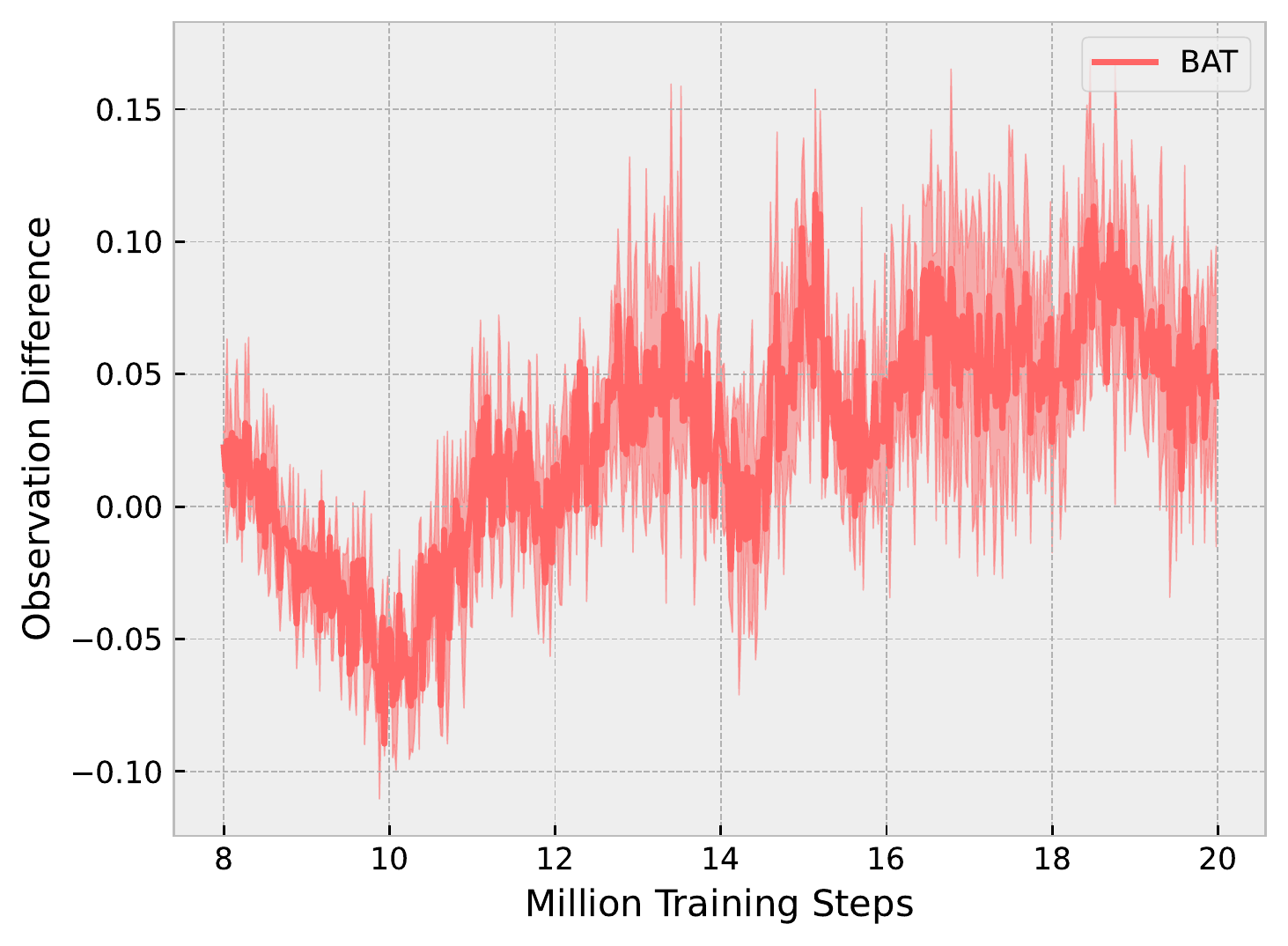}\label{mmm0_obs_diff}}
	\subfigure[\textit{MMM9}]{\includegraphics[width=.32\textwidth, height=4.25cm]{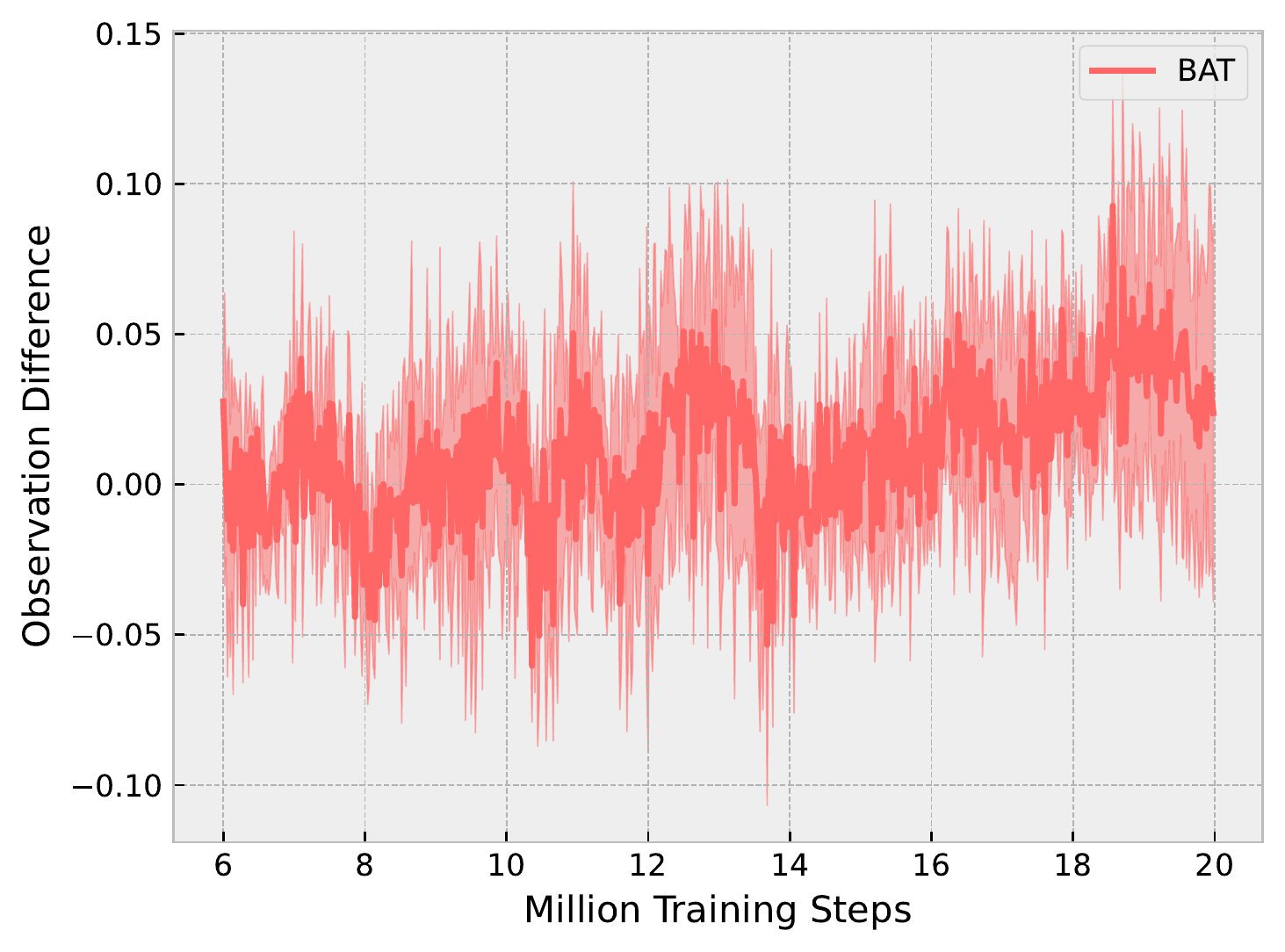}\label{mmm9_obs_diff}}
	\subfigure[\textit{MMM3}]{\includegraphics[width=.32\textwidth, height=4.25cm]{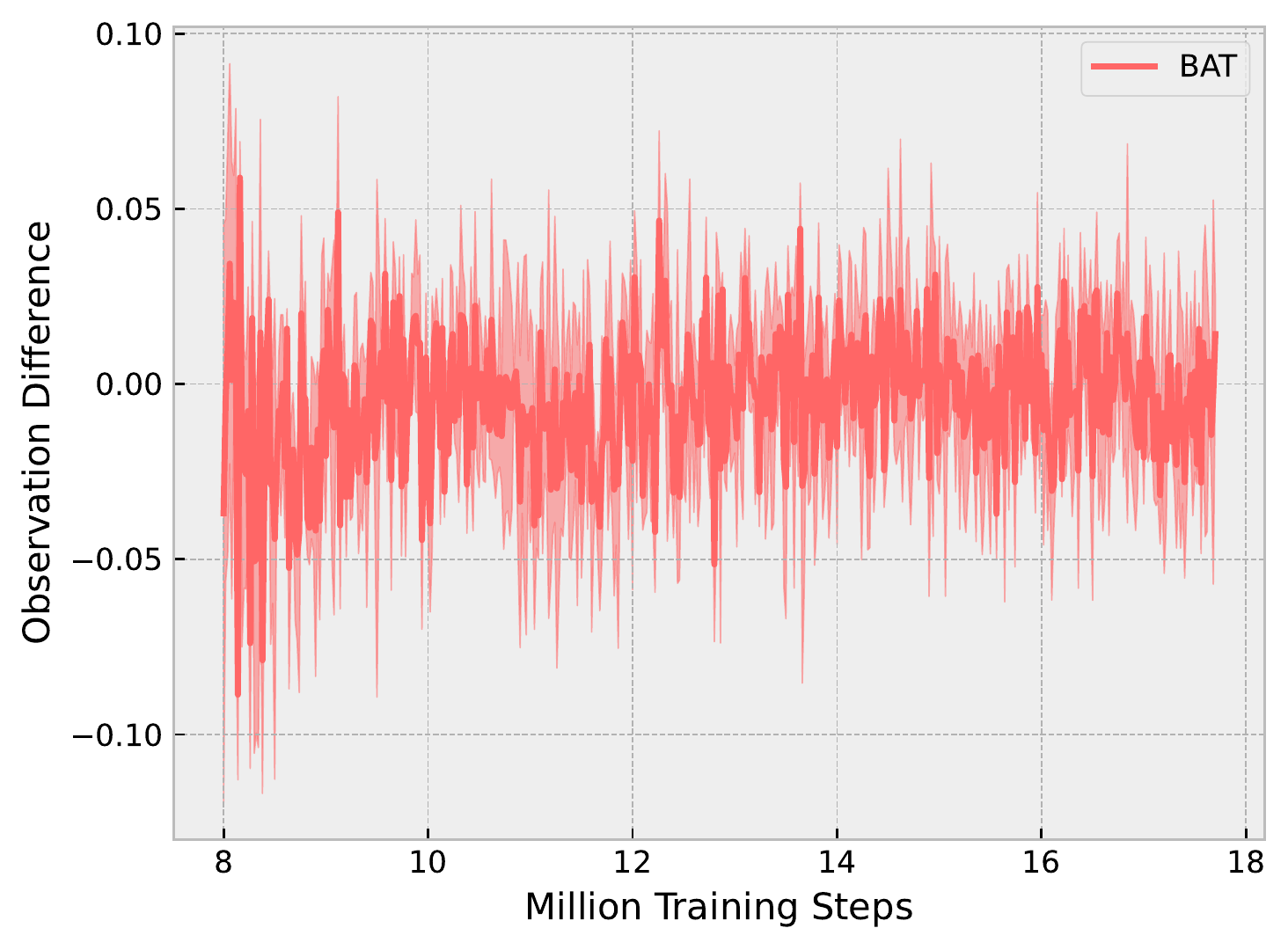}\label{mmm3_obs_diff}}
	
	\caption{The observation difference during the training.}
	\label{fig_obs_diff}
\end{figure*}

In this part, we present the experiment results about the adversary agent's influences on its teammates. We first show the influences on teammates' observations and then show the resulting influences on teammates' behaviors. Specifically, during the training, we compute the encoding vector distance of teammates' observations, i.e., $\|\mathcal{E}(o;\theta)-\bar{\mathcal{E}}(o)\|_2$, for both benign and abnormal episodes. For benign episodes, we denote the mean of the distances for all time steps as $\mathcal{L}_{obs}$. For abnormal episodes, we denote the mean of distances as $r_{obs}$. 
Note that although we minimize $\mathcal{L}_{obs}$ for benign episodes, it is still not zero. Therefore, $\mathcal{L}_{obs}$ serves as a baseline and we use $diff_{obs}=r_{obs}-\mathcal{L}_{obs}$ to represent the observation difference between benign episodes and abnormal episodes. This value can reflect the influence of the adversary agent on teammates' observations. We show the value of $diff_{obs}$ throughout the training in Figure \ref{fig_obs_diff}. From this figure, we can see that the value of $diff_{obs}$ in the \textit{3s\_vs\_3z}, \textit{8m}, \textit{MMM0}, and \textit{MMM9} scenarios all show an increasing trend. However, the increasing may not be that significant because the number of trigger actions we perform is very small. In comparison, the value of $diff_{obs}$ in the \textit{MMM3} scenario only fluctuates around zero, which indicates that the adversary agent does not affect its teammates' observations effectively.

Besides the influence on observations, we are also curious about whether the teammates' behavior are affected given their affected observations. To evaluate that, we consider three settings: 1) we equip the whole team with the benign policy $\bm{\pi}_b$ trained during the benign training period; 2) we equip the teammates with $\bm{\pi}_b$ while equipping the adversary agent with the backdoored policy that embeds both benign and abnormal behaviors; 3) we equip the whole team with the backdoored policies (note that teammates and the adversary agent do not share the same backdoored policy because the adversary is in charge of performing trigger actions). The first setting represents the benign behavior, denoted as \textit{Benign}. The second setting will indicate that when the adversary agent performs trigger actions, how the original benign team policy will react, denoted as \textit{Mixed\_team}. The third setting is the backdoor attack setting, denoted as \textit{Backdoor}. The comparison between the second and third setting can show that whether our backdoor attack training can enable the adversary agent to change teammates' behaviors based on trigger actions. We measure the behavior based on the cumulative task return because agents' behavior can directly affect the task reward received in every step. We test the three settings in both benign and abnormal mode. The number of test episodes are $128$ for each mode. The results of cumulative return and trigger times are shown in Table \ref{t1} and \ref{t2}. From Table \ref{t1}, we can see that in benign episodes, all settings get a high return and complete the task well. However, for abnormal episodes (Figure \ref{t2}), the cumulative returns of the \textit{Mixed\_team} setting are very different from that of the \textit{Backdoor} setting. This demonstrate that the original benign team policy will not react much to the trigger actions performed by the adversary agent and the teammates still try to complete the task. However, in the \textit{Backdoor} setting, after the adversary agent performs the trigger actions, the teammates' behavior will also change, which results in the low returns. These results indicate that our backdoor training can implant malicious behaviors in the teammates' policies and the adversary agent can trigger the malicious behaviors by trigger actions. 

\begin{table}[!ht]
	\centering
	\caption{Performance in benign episodes}
	\begin{tabular}{cccc}
		\hline
		Scenario/Setting & \textit{Benign} & \textit{Mixed\_team} & \textit{Backdoor}  \\ \hline
		\textit{3s\_vs\_3z}  & $18.65\pm2.99$,   & $18.59\pm3.04$   & $19.00\pm2.63$   \\ 
		 \textit{8m}    & $19.1\pm2.38$     & $19.1\pm2.38$ & $19.1\pm2.38$       \\ 
		 \textit{MMM0}    & $19.05\pm2.90$    & $19.16\pm2.77$ &  $18.98\pm2.79$       \\ 
		 \textit{MMM9}    & $19.82\pm3.32$     & $18.60\pm3.43$ &  $19.78\pm2.93$                \\ \hline
	\end{tabular}\label{t1}
\end{table}

\begin{table}[!ht]
	\centering
	\caption{Performance in abnormal episodes}
	\begin{tabular}{cccc}
		\hline
		Scenario/Setting & \textit{Benign} & \textit{Mixed\_team} & \textit{Backdoor}  \\ \hline
		\textit{3s\_vs\_3z}  & $18.71\pm2.82$     & $6.88\pm2.17$ &  $0.05\pm0.18$       \\ 
		\textit{8m}    & $18.65\pm2.9$    & $8.90\pm1.69$ &  $0.66\pm0.58$       \\ 
		\textit{MMM0}    & $19.20\pm2.84$     & $7.21\pm3.16$    &  $0.77\pm0.73$       \\ 
		\textit{MMM9}    & $19.18\pm2.81$     &  $12.26\pm4.24$    &  $1.11\pm1.20$          \\ \hline
	\end{tabular}\label{t2}
\end{table}

\end{document}